\documentclass[twocolumn]{aastex7}

\usepackage{pgfplotstable}
\usepackage{booktabs}      
\usepackage{xcolor}

\newcommand{\mgii}{Mg\,{\sc ii}}

\newcommand{\siii}{Si\,{\sc ii}}

\newcommand{\oiii}{O\,{\sc iii}}
\newcommand{\hbeta}{H-$\beta$}
\newcommand{\halpha}{H-$\alpha$}

\newcommand{\civ}{C\;{\sc iv}}
\newcommand{\cii}{C\;{\sc ii}}

\newcommand{\feii}{Fe\;{\sc ii}}
\newcommand{\oi}{O\;{\sc i}}
\newcommand{\hi}{H\;{\sc i}}
\newcommand{\hii}{H\;{\sc ii}}
\newcommand{\siiv}{Si\;{\sc iv}}

\newcommand{\kms}{$\rm{km s^{-1}}$}

\begin{document}
\title{EIGER VIII: First stars signatures in the connection between OI absorption and Galaxies in the Epoch of Reionization}

\correspondingauthor{Rongmon Bordoloi}
\email{rbordol@ncsu.edu}\tabletypesize{\footnotesize}

\author[0009-0005-8793-2445]{Jack Higginson}\email{jbhiggi3@ncsu.edu}
\affiliation{Department of Physics and Astronomy, North Carolina State University, Raleigh, 27695, North Carolina, USA}

\author[0000-0002-3120-7173]{Rongmon Bordoloi}\email{rbordol@ncsu.edu}
\affiliation{Department of Physics and Astronomy, North Carolina State University, Raleigh, 27695, North Carolina, USA}

\author[0000-0003-3769-9559]{Robert A.~Simcoe}\email{simcoe@space.mit.edu}
\affiliation{MIT Kavli Institute for Astrophysics and Space Research, 77 Massachusetts Ave., Cambridge, MA 02139, USA}

\author[0000-0003-2871-127X]{Jorryt Matthee}\email{jorryt.matthee@ista.ac.at}\affiliation{Institute of Science and Technology Austria (ISTA), Am Campus 1, 3400 Klosterneuburg, Austria} 

\author[0000-0001-9044-1747]{Daichi Kashino}\email{kashinod.astro@gmail.com}
\affiliation{National Astronomical Observatory of Japan (NAOJ), 2-21-1, Osawa, Mitaka, Tokyo 181-8588, Japan}

\author[0000-0003-0417-385X]{Ruari Mackenzie}\email{mruari@ethz.ch}
\affiliation{Department of Physics, ETH Z{\"u}rich, Wolfgang-Pauli-Strasse 27, Z{\"u}rich, 8093, Switzerland}

\author[0000-0001-5346-6048]{Ivan Kramarenko}\email{im.kramarenko@gmail.com}\affiliation{Institute of Science and Technology Austria (ISTA), Am Campus 1, 3400 Klosterneuburg, Austria} 

\author[0000-0002-6423-3597]{Simon J.~Lilly}\email{simonlilly2024@gmail.com}
\affiliation{Department of Physics, ETH Z{\"u}rich, Wolfgang-Pauli-Strasse 27, Z{\"u}rich, 8093, Switzerland}

\author[0000-0003-2895-6218]{Anna-Christina Eilers}\email{eilers@mit.edu}
\affiliation{MIT Kavli Institute for Astrophysics and Space Research, 77 Massachusetts Ave., Cambridge, MA 02139, USA}

\author[0000-0003-3997-5705]{Rohan~P.~Naidu}\email{rnaidu@mit.edu}
\altaffiliation{NASA Hubble Fellow}
\affiliation{MIT Kavli Institute for Astrophysics and Space Research, 77 Massachusetts Ave., Cambridge, MA 02139, USA}

\author[0000-0002-5367-8021]{Minghao Yue}\email{myue@mit.edu}
\affiliation{MIT Kavli Institute for Astrophysics and Space Research, 77 Massachusetts Ave., Cambridge, MA 02139, USA}

\begin{abstract}

We investigate the association between galaxies and neutral \oi\ absorption systems at $z\sim6$, which trace metal-enriched gas during the epoch of reionization. We identify 40 galaxies across six quasar fields, residing in 15 overdensities within 300~kpc of the background sightlines. Five \oi\ absorption systems are associated with five of these overdensities, yielding a covering fraction of $0.27^{+0.13}_{-0.10}$ within 300~kpc. The absorption occurs beyond typical virial radii, indicating that the gas traces extended overdensity environments rather than individual galaxy halos, unlike the $z\sim0$ CGM which is largely bound to halos. These galaxy-associated absorbers account for $\sim35\%$ of all \oi\ systems seen in blind quasar surveys, implying the remainder arise in lower-mass galaxies below our detection threshold or in dense neutral IGM pockets. The CGM around these galaxies contains $\gtrsim 2\times10^6~M_{\odot}$ of oxygen, comparable to the ISM oxygen mass of the galaxies themselves, suggesting that the surrounding environment holds as much metal mass as the galaxies. All five galaxy-associated systems show significantly higher $\log(N_{\rm CII}/N_{\rm OI})$ ratios than absorbers lacking galaxy associations. Furthermore, relative abundance ratios ([Si/O], [C/O]) reveal that four of the five exhibit enrichment patterns consistent with Population~III nucleosynthesis at the outskirts of galaxy overdensities. These rare systems offer a unique window into the role of first-generation stars in shaping the early metal enrichment of galaxies and their environments.
\end{abstract}


\section{Introduction}

The first generations of stars formed shortly after the Epoch of Recombination within dark matter overdensities, producing the first significant quantities of metals and enriching the intergalactic medium (IGM) while reionizing hydrogen and helium \citep[e.g.,][]{klessen_2023_annual_review}. These metals and gas eventually coalesced into the first galaxies, which over cosmic time evolved into Milky Way–like systems today. Studying these young galaxies and their environments provides critical insight into galaxy formation and evolution, constraining models of stellar feedback, gas accretion, and chemical enrichment \citep{Faucher_2023, Peroux_Nelson_2024}.  

At low redshift, decades of observations using deep ground-based spectroscopy and HST data of background quasars and foreground galaxies have built a rich understanding of the circumgalactic medium (CGM) and its interplay with galaxies \citep[e.g.,][]{tumlinson2017}. The CGM plays a crucial role in gas recycling, star formation regulation, and quenching; understanding its structure and composition informs galaxy growth models \citep{Peng_2015, Zinger_2020}.

High-redshift studies, leveraging ground- and space-based NIR spectroscopy of quasars, reveal that the early Universe was already chemically enriched by the epoch of reionization \citep[e.g.,][]{Davies_2023a, Cooper_2019, christensen_metal_2023}. Among tracers of neutral and low-ionization gas, \oi\ has emerged as a key probe: surveys report an increase in \oi\ number density at $z \gtrsim 5-6$ \citep{Becker_2019, christensen_metal_2023, sebastian2023, Sodini_2024}, while higher-ionization species like \civ\ and \siiv\ decline, consistent with a softer and evolving UV background at the end of reionization \citep{Davies_2023a, Davies_2023b, Dordorico_2022, Cooper_2019}.

Historically, connecting absorption systems with their host galaxies was challenging due to the difficulty of obtaining deep, high-redshift galaxy spectroscopy. This has changed with the advent of JWST, providing unprecedented views of early galaxies and their impact on surrounding gas through surveys such as EIGER \citep{Kashino2023} and ASPIRE \citep{ASPIRE_Wang_2023}. First systematic studies of cool, low-ionization CGM traced by \mgii\ around $z\sim6$ galaxies suggest that high-redshift CGM differs significantly from its low-redshift counterpart: these galaxies typically have low masses, and strong stellar feedback can eject metals far beyond the halo's gravitational potential \citep{bordoloi_eigerIV_2024}. Such pioneering studies now allow systematic investigations of the CGM in the early Universe.  

The UV background during reionization was inhomogeneous and rapidly evolving \citep[e.g.,][]{furlanetto_2005, Oppenheimer_2009, Finlator_2013, Becker_Bolton_Lidz_2015, Bosman_2022}, initially dominated by Population III stars \citep{Nakajima_2022, Nakajima_2025}, then by early star-forming galaxies \citep{Jiang_2022, Matthee_2022}, with AGN contributing significantly by $z \sim 5$ \citep{Daloisio_2017, Kulkarni_2019, Giguere_2020}. Modeling indicates that pockets of Population III star formation may persist throughout the Epoch of Reionization \citep{Tornatore_2007, Mebane_2018, Johnson_2019, Venditti_2023, Zier_2025}.  

The large reservoir of neutral \oi\ observed in quasar absorption lines at high redshift is typically self-shielded, providing robust estimates of chemical abundances. Studying the association of these absorbers with galaxies offers unique insight into the CGM of early galaxies, and into potential contributions from Population III stars to early chemical enrichment.  

In this work, we study the cool, neutral CGM around galaxies at $z \sim 6$, traced by \oi\ absorption from the EIGER survey (\textit{Emission-line Galaxies and Intergalactic Gas in the Epoch of Reionization}; \citealt{Kashino2023}). We focus on \oi\ absorbers associated with [\oiii]$_{5008}$-emitting galaxies at $5.875 < z < 6.378$ within 300 kpc of six background quasars. We quantify absorber incidence, column densities, and mass estimates, and analyze ionic abundances to probe chemical enrichment, including potential contributions from Population~III stars.

This paper is organized as follows. Section 2 describes the observations; Section 3 details our methods for galaxy selection, redshift estimation, galaxy properties, defining galaxy overdensities, absorber measurements, and completeness; Section 4 presents the results, including the association of \oi\ absorption with galaxies, the incidence of galaxy-associated \oi\ absorbers, their chemical abundances, and physical interpretations; Section 5 summarizes our conclusions. Throughout, we adopt a $\Lambda$CDM cosmology with $H_0 = 67.66~\mathrm{km~s^{-1}~Mpc^{-1}}$ \citep{Planck_Collaboration_2020}.

\section{Observations} \label{sec:observations}

 EIGER (Program ID 1243; PI: S.~J.~Lilly) is a large JWST Wide Field Slitless Spectroscopy (WFSS) program totaling 126.5 hours, conducting deep WFSS observations in six extragalactic fields, each centered on an ultra-luminous quasar at $z \geq 6$. These quasars have also been observed with deep optical and near-infrared (NIR) medium- and high-resolution spectroscopy with Magellan/FIRE, VLT/X-shooter, and Keck HIRES/MOSFIRE, providing over 100 hours of complementary NIR spectroscopy across the six sightlines. We refer the reader to \citet{Eilers2023} and \citet{durovica2025} for data reduction and details of the ground-based quasar spectroscopy. In addition, all fields are covered by deep JWST/NIRCam and HST imaging. A detailed description of the survey design, methodology and spectroscopic and photometric observation details are provided in \citet{Kashino2023}, \citet{EIGERII}, and \citet{bordoloi_eigerIV_2024}. In this work, we make use of the complete EIGER survey of $z \geq 5.9$ galaxies spanning all six extragalactic fields \citep{EIGERVII}. 

We note that all six quasars were chosen with prior knowledge of metal absorption systems along their lines of sight, which could in principle bias any attempt to associate galaxies with those absorbers. In contrast, a galaxy-centric approach starts with quasar–galaxy pairs without prior knowledge of absorption and then builds statistics systematically \citep[see e.g.,][]{Bordoloi2011,Werk2013,Bordoloi_COS_2014}. However, once a blind, well-characterized galaxy survey is in place, the two approaches converge. This is conceptually similar to a program that combines quasar spectroscopy performed without prior absorber selection with a complete galaxy survey. This work employs deep blind galaxy searches in each of the EIGER fields, yielding an emission-line complete sample of spectroscopic galaxies \citep{Matthee_2022,bordoloi_eigerIV_2024} and a complete census of absorption lines along these sightlines (Section~\ref{sec:completeness}). By combining these two elements—a galaxy survey with quantified completeness and an absorption survey with well-defined sensitivity—we establish a robust framework for conducting a census of CGM absorption around galaxies. This combined approach has been successfully demonstrated at low redshifts \citep[][]{Burchett_2019}.

\begin{figure}[]
\centering
\includegraphics[width=1.0\linewidth]{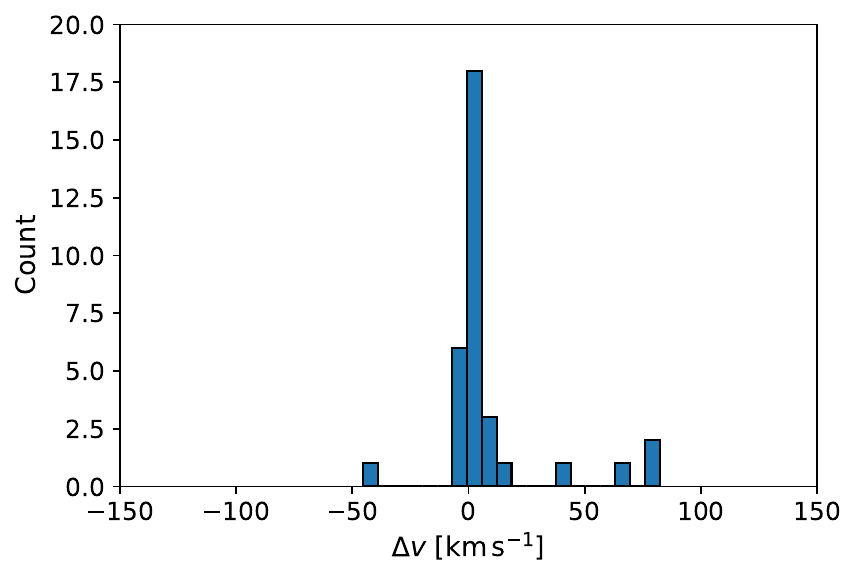}
\caption{\label{fig:delta_hist} 
Redshift uncertainty distribution of the [\oiii] emitter EIGER galaxies. The histogram shows the difference in redshift between emission-line measurements from NIRCam Grism Module A and Module B spectra, expressed in \kms. Individual redshifts were derived from Gaussian fits to the [\oiii] and \hbeta\ lines. The two modules yield consistent results with a mean offset and standard deviation of $\Delta z = 0.0028 \pm 0.0073$ ($\Delta v = 9.3 \pm 24$ \kms).
}

\end{figure}

\section{Methods}\label{sec:methods}
In this section we describe how galaxy redshifts and properties are measured and how absorption lines are identified and measured. 

\begin{figure*}[t]
\centering
\includegraphics[width=1.0\linewidth]{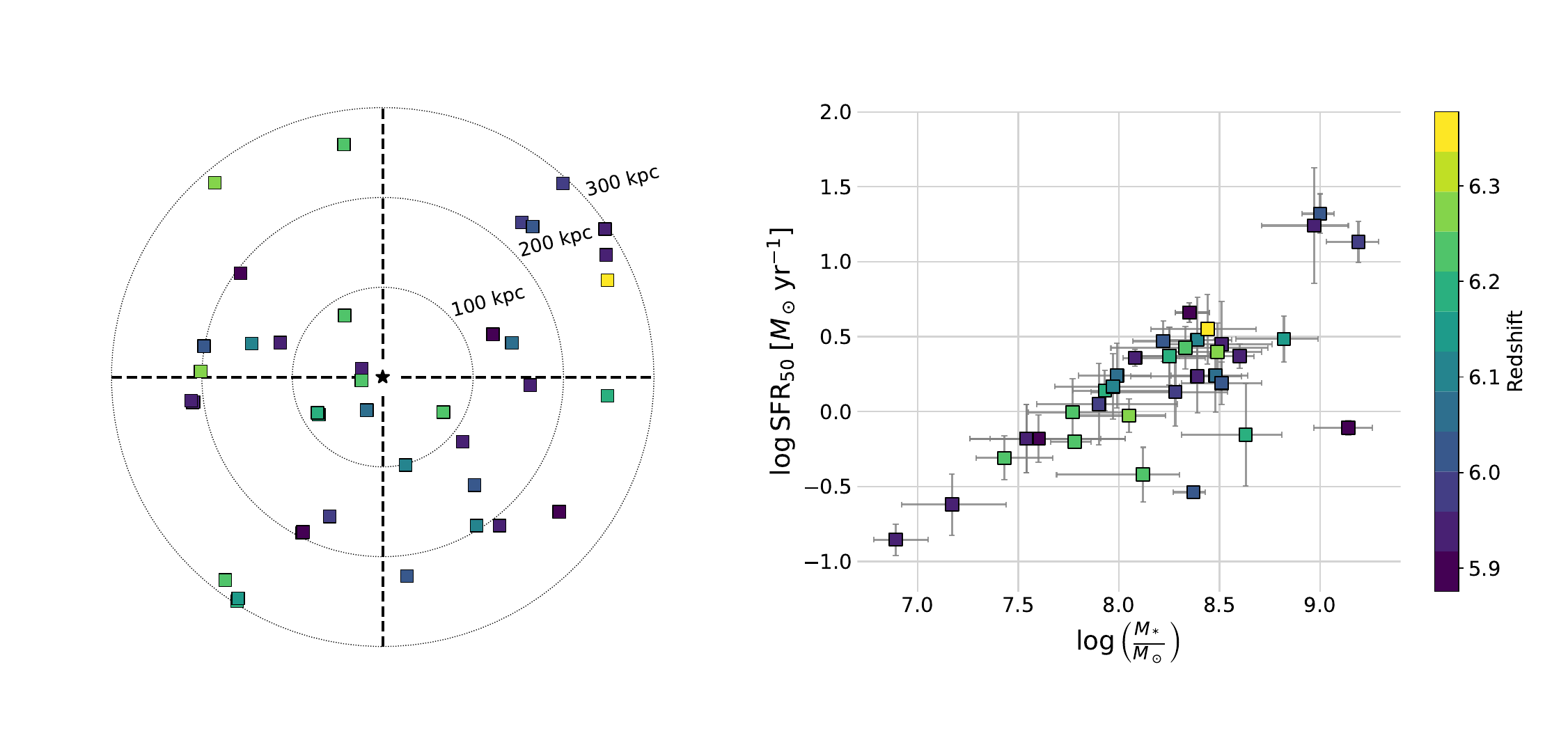}
\caption{\label{fig:bullseye_SFR_Mstar} 
Properties of 40 spectroscopically confirmed galaxies at $5.9 \leq z \leq 6.4$ within 300 kpc of background quasars in the EIGER survey. Each square denotes a galaxy and is color-coded by redshift. \textit{Left:} Angular positions and impact parameters relative to the background quasar (black star), with dashed concentric rings marking 100, 200, and 300 kpc at $z=6$. \textit{Right:} Stellar mass versus star-formation rate, averaged over the last 50 Myr. Error bars show the 16th–84th percentile uncertainties in stellar mass and SFR.
}

\end{figure*}

\subsection{Galaxy Redshifts}

Since our focus is on galaxies at $z \sim 6$ near the quasar sightlines, we begin with the full EIGER catalog of [\oiii] emitters \citep{EIGERVII}, which contains 948 galaxies detected via [\oiii] emission within $5.33 \leq z \leq 6.97$. From this catalog we select galaxies that (1) lie at least 10,000 \kms\ in the foreground of each quasar $c\,\frac{z_{\rm QSO}-z_{\rm gal}}{1+z_{\rm QSO}} > 10{,}000~{\rm km\,s^{-1}}$, (2) are within a projected impact parameter of $300~\mathrm{pkpc}$, and (3) have \oi\ $\lambda 1302$ coverage in the quasar NIR spectra outside of the Gunn–Peterson trough. This yields a sample of 40 galaxies. Of these, 33 fall within the central $\approx 4.6~{\rm arcmin}^2$ of the EIGER spectroscopic footprint \citep{bordoloi_eigerIV_2024}, observed with both JWST/NIRCam reverse grism modules (A and B). Because the two modules disperse light in opposite directions, they provide independent moderate-resolution ($R \sim 1500$) spectra of the same target, enabling us to assess redshift measurement uncertainties.

Following the procedure in \citet{bordoloi_eigerIV_2024}, we visually inspect each NIRCam 2D spectrum using the custom Python API \texttt{zgui} \citep{rbcodes}, extract 1D spectra from each module, and fit Gaussians to the $[O_{III}]_{5008}$ emission line to determine redshifts.  Redshifts from the two modules are consistent to a mean velocity offset of $\Delta v = 9.3$ \kms\  with a standard deviation of $24$ \kms\  (Figure~\ref{fig:delta_hist}). We therefore adopt an average redshift uncertainty of $\approx 24$ \kms\ for the [\oiii] emitters, sufficient for associating metal absorption systems with their host galaxies. Hereafter, we focus on this sample of 40 galaxies.

\subsection{Galaxy stellar mass, star-formation rate, halo mass}

To estimate galaxy stellar masses and star-formation rates (SFRs), we follow the procedure of \citet{Matthee_2022}. We first measure the [\oiii] and \hbeta\ emission-line fluxes of each galaxy, and combine these with JWST/NIRCam broad-band photometry (F115W, F200W, and F356W) to perform spectral energy distribution (SED) fitting at the galaxy redshift using the \texttt{Prospector} code \citep{Johnson2021}. We assume a \citet{Chabrier2003} initial mass function (IMF), adopt MIST isochrones \citep{Choi2016,Dotter2016}, a non-parametric star-formation history model \citep{Naidu_SF_model}, and the dust attenuation curve of \citet{Calzetti2000}. Galaxy SFRs are derived by averaging over the most recent $50$ Myr of star-formation history from the \texttt{Prospector} posteriors.

Figure~\ref{fig:bullseye_SFR_Mstar} (left) shows the impact parameter distribution of the 40 EIGER galaxies relative to the background quasars (black star symbols). The galaxies are randomly distributed with respect to the quasar sightlines. The right panel shows their stellar masses and SFRs, with symbols color-coded by galaxy redshift. Following \citet{bordoloi_eigerIV_2024}, we estimate halo masses and virial radii ($R_{\rm vir}$) using abundance matching from \citet{Behroozi_2019}. Uncertainties in $R_{\rm vir}$ are propagated from the $16^{\rm th}$ and $84^{\rm th}$ percentile stellar mass errors, giving a mean uncertainty of $\Delta R_{\rm vir} = 1.8~\mathrm{kpc}$. The galaxies span stellar masses of $6.8 \leq \log (M_*/M_{\odot}) \leq 9.3$ and star-formation rates of $-1 \leq \log (\mathrm{SFR}{50}/M{_\odot}\,\mathrm{yr}^{-1}) \leq 1.5$, with all systems being star-forming by selection.

One galaxy (z=5.910384, galaxyID=15362, qsoID = J159) has an exceptionally high stellar mass of $\log (M_*/M_{\odot}) = 11.05$, beyond the calibrated range of the \citet{Behroozi_2019} relation. This object exhibits a compact, point-like morphology and likely hosts an obscured AGN, placing it among the recently identified population of Little Red Dots (LRDs) \citep[e.g.,][]{Matthee2024LRD}. We do not detect broad \hbeta\ for this object. However, this is not unexpected, as LRD can have \halpha\ / \hbeta\ values as high as $\sim 10$ (\cite{Brooks_2025}, and Torralba et al., in prep). As robust AGN modeling at these redshifts is not yet established, we estimated its halo mass by linearly extrapolating the \citet{Behroozi_2019} relation from the stable stellar mass regime. We explicitly note where this assumption may affect our results in the following sections. Additionally, this galaxy has not been plotted in the right panel of \ref{fig:bullseye_SFR_Mstar}.

\subsection{Galaxy Overdensities}
\label{sec:method_overdensities}

A significant fraction of $z \geq 6$ galaxies in the EIGER survey are not isolated but are found in overdense environments, either as mergers or members of ``proto-groups" \citep[see, e.g.,][]{EIGERVII}. The EIGER [\oiii] galaxy catalog already merges individual [\oiii] emission clumps within $2\arcsec$ into a single system, even when they exhibit complex morphologies or merger signatures. To identify larger-scale overdensities, we therefore search for galaxies that lie at similar redshifts but are widely separated on the sky.

Within $300~\mathrm{kpc}$ from the QSO sightlines, we identify overdensities around the 40 selected EIGER galaxies using a friends-of-friends algorithm \citep{rbcodes}. Galaxies are grouped as an overdensity if they lie within $1~\mathrm{pMpc}$ in projected separation and within $\pm 1000$ \kms\ in redshift space. A galaxy is considered part of an overdensity only if all members are ``fully" connected, i.e., each galaxy in the overdensity satisfies these conditions with all other members. Around each detected \oi\ absorption system, we further search for overdensities within $\pm$1500 \kms\ of the absorber redshift, and merge these associations to ensure all clustered galaxies are treated as a single overdensity. This conservative approach ensures that we capture the full galaxy environment of \oi\ absorbers. The 40 galaxies studied in this work are part of 15 distinct overdensities with each grouping containing 1 -- 5 members. The number of galaxies in each grouping is reported in Table~\ref{tab:logN_table}. The typical velocity dispersion of these overdensities is $449.2~\mathrm{kms^{-1}}$. 

Throughout this work, we define the galaxy with the smallest impact parameter to the background quasar sightline within each galaxy overdensity as the primary member. This convention is adopted solely to simplify the presentation and does not imply a unique physical association between the intervening absorption system and this galaxy. In fact, the concept of a unique absorber–galaxy association at high redshift is significantly less robust than in the low-redshift Universe (e.g., \citealt{bordoloi_eigerIV_2024}). As discussed in Section~\ref{sec:variation_OI_around_galaxies}, none of the \oi\ absorption systems lie within the virial radius of any individual halo. Even at low redshift, the galaxy with the smallest impact parameter is not necessarily the one with the smallest velocity offset from the absorber, and therefore does not constitute evidence for a unique absorber–galaxy pair \citep{Hamanowicz_2020}. This behavior is also observed for galaxies residing in groups \citep{Bordoloi2011}. Our analysis therefore focuses on the broader environment, which we argue is most appropriately characterized by galaxy overdensities, and our choice of a primary galaxy does not affect our main conclusions.

\subsection{Absorption line detection and measurements}
\label{sec:methods_line_detection}

We follow the procedures outlined in \citet{bordoloi_eigerIV_2024} to identify and measure the strength of \oi\ absorption systems around EIGER galaxies, and provide a brief summary here. We first use a custom Python-based 1D spectrum viewer to identify and tabulate all intervening absorption lines in the EIGER quasar spectra. A semi-automated CGM absorption-line analysis pipeline is then used to perform the measurements. For each galaxy, we extract a $\pm1500$ \kms\ spectral window centered on its systemic redshift, for all transitions of interest. In this work, we analyze slices around \oi\ $\lambda1302$, \siii\ $\lambda1260$, \siii\ $\lambda1304$, \siii\ $\lambda1527$, \cii\ $\lambda1334$, \siiv\ $\lambda1393$, \siiv\ $\lambda1402$, \civ\ $\lambda1548$, and \civ\ $\lambda1550$. Each slice is locally continuum-normalized using an automated Legendre polynomial fit, with the polynomial order chosen via the Bayesian Information Criterion. The continuum fits and absorption features are visually inspected, and both rest-frame equivalent widths ($W_{r}$) and apparent optical depth (AOD) column densities are measured. We define a detection as any absorption system with $W_{r} \geq 3\sigma_{w}$, where $\sigma_{w}$ is the $1\sigma$ uncertainty in $W_{r}$. We note that this uncertainty contains both the statistical uncertainty due to signal-to-noise ratio of the data and the continuum fitting uncertainty. Non-detections are reported as $2\sigma_{w}$ upper limits measured within $\pm$100 \kms\ of the systemic redshift of the galaxy. For column density we report the $2\sigma$ uncertainty on AOD column density as upper limit on non detections. When a transition is detected in multiple instruments, each spectrum is analyzed independently, and the inverse-variance–weighted mean is adopted as the final $W_{r}$. The details of this automated approach and full documentation of the APIs used can be found in \cite{rbcodes}.

For all detections, we perform Voigt profile fitting using the \texttt{rbvfit} package, which employs Markov Chain Monte Carlo (MCMC) sampling to simultaneously fit multiple absorption systems across different instruments \citep{rbvfit}. This approach yields robust column densities, resolves the kinematic structure, and enables accurate modeling of blended lines. Additionally, since we obtain marginalized posterior distribution for the fitted parameters, the results yield accurate column density estimates even if doppler $b$ parameter or velocity centroids are not well constrained in moderate resolution spectra (see discussion in \citealt{bordoloi_eigerIV_2024}). In our case, most of the \oi\, \siii\, \cii\ transitions are all covered by higher resolution HIRES spectroscopy and thus their properties are well constrained. \texttt{rbvfit} supports two MCMC samplers, \texttt{emcee} \citep{emcee} and \texttt{zeus} \citep{ZEUS1,ZEUS2}, both of which are used in this work depending on the complexity of the fit. We typically adopt 50 walkers with initial perturbations of either $10^{-3}$ or $10^{-6}$, and default chain lengths of $8000$–$20{,}000$ steps, ensuring the total number of steps exceeds 50 times the autocorrelation time. Additional steps are run as needed to achieve convergence. In each case, a flat prior is assumed with reasonable physical bounds.

\begin{figure*}[t]
\centering
\includegraphics[width=\linewidth]{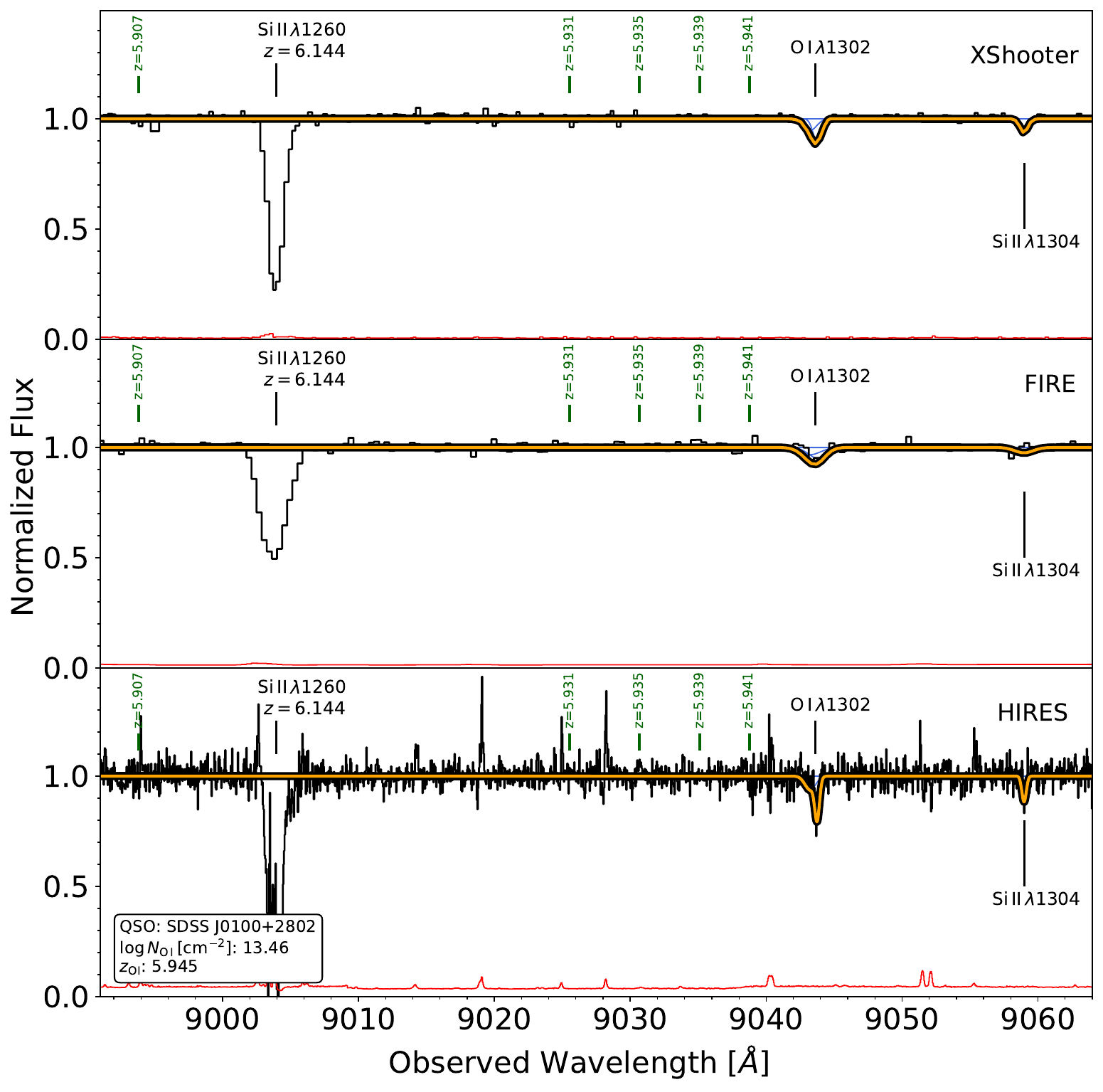}
\caption{
Voigt profile fits to \oi\ and \siii\ absorption profiles at $z=5.945$ toward the quasar SDSS J0100+2802. Shown are the flux (black), error spectrum (red), best-fit Voigt model (orange/black), and individual Voigt components (thin blue), using data from FIRE, X-shooter, and HIRES. Vertical green lines mark galaxy members of the associated overdensity. Spectra from all instruments and all transitions are simultaneously fit using the \texttt{rbvfit} Bayesian absorption line fitter.
}
\label{fig:voigt_subplot_J0100_7406_XShooter_FIRE_HIRES_OI}
\end{figure*}

\begin{figure*}[h]
\centering
\includegraphics[width=1.0\linewidth]{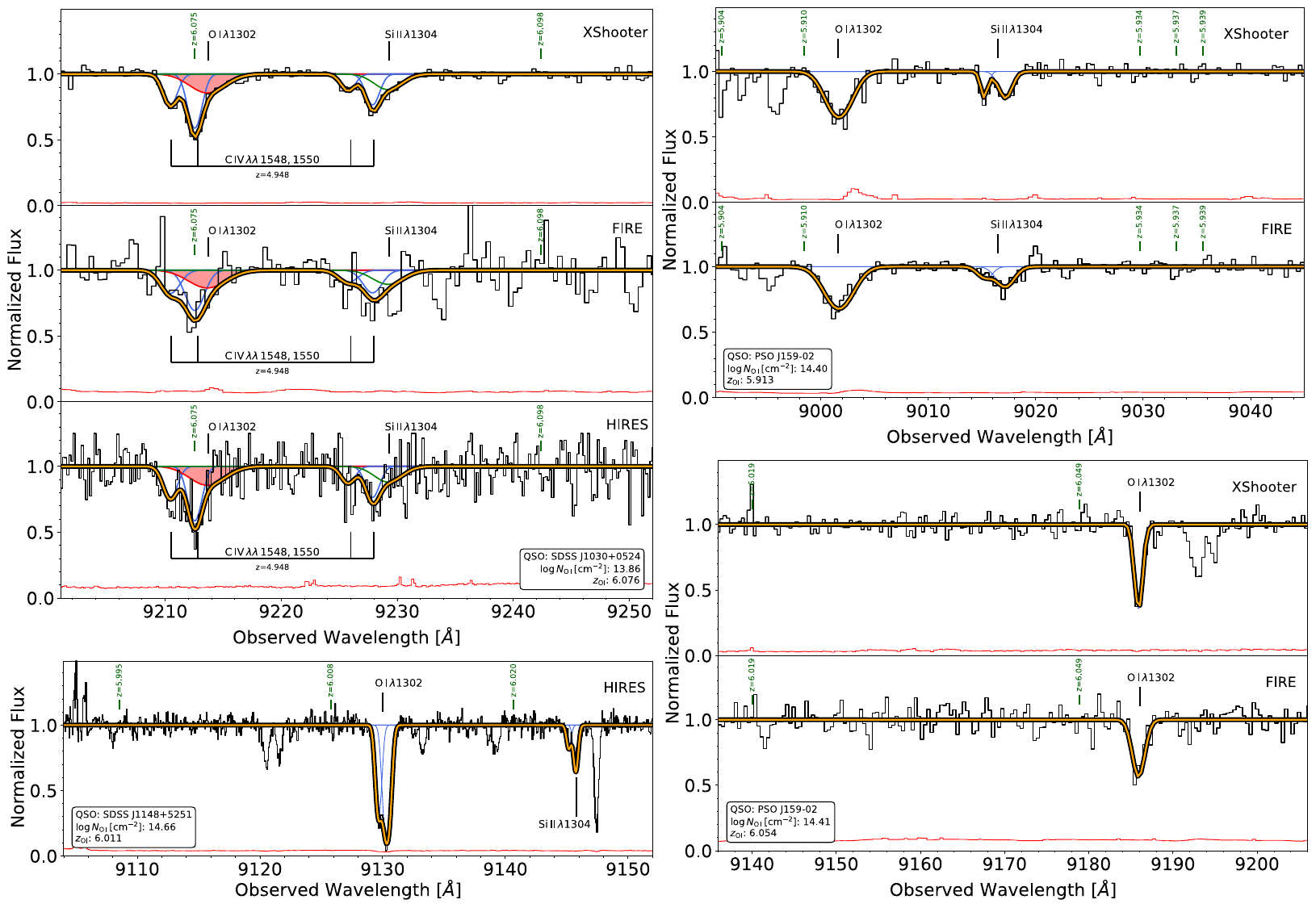}
\caption{\label{fig:all_fits} Voigt profile fits: Flux (black), error (red), most likely Voigt fit (orange and black). The individual Voigt component fits are shown for all instruments available for the quasar sightline. Lines and text present to specify the velocity centroids for each element. The vertical green lines indicate the galaxy members of the overdensity. \textit{Top left}: A Voigt profile fit for \oi\ and $Si\,\textsc{ii}\,\lambda1304$ at $z=6.075$, and $C\,\textsc{iv}\,\lambda\lambda1548,1550$ at $z=4.9484$ around the quasar SDSS J1030+0524. The HIRES data has been re-binned and clipped for aesthetic purposed. The colors represent different transitions: blue for \civ, green for \siii, and red for \oi, with \oi\ being filled in red. \textit{Top right}: A Voigt profile fit for \oi\ and $Si\,\textsc{ii}\,\lambda1304$ at $z=6.011$ around the quasar SDSS J1148+5251. \textit{Bottom left}: A Voigt profile fit for \oi\ and $Si\,\textsc{ii}\,\lambda1304$ at $z=5.913$ around the quasar PSO J159–02. \textit{Bottom right}: A Voigt profile fit for \oi\ at $z=6.054$ around the quasar PSO J159–02.}
\end{figure*}

Figure~\ref{fig:voigt_subplot_J0100_7406_XShooter_FIRE_HIRES_OI} shows an example of Voigt profile fits to \oi\ $\lambda1302$ and \siii\ $\lambda1304$ lines simultaneously modeled across FIRE, X-shooter, and HIRES. We fit a two component Voigt profile with a mean  velocity offset from the host galaxy of $v_{offset}=$ 157 \kms\ .  This joint fitting maximizes information from all instruments while preserving the kinematic detail provided by higher-resolution data. Galaxy overdensities associated with this system are marked with green vertical ticks. This system has a total \oi\ column density $\log N_{OI}/cm^{-2} =$ 13.46$^{+0.11}_{-0.15}$. All other \oi\ fits are presented in Figure \ref{fig:all_fits}. All absorption line measurements are presented in Table \ref{tab:logN_table}. 

We note that an \oi\ and \siii\ absorption system is detected along the J1030 quasar sightline at $z \sim 6.075$, blended with an intervening \civ\ absorber at $z \sim 4.948$ (Figure~\ref{fig:all_fits}, top left panels; see also \citealt{Diaz_2021}). The $z \sim 6.076$ system is confirmed by the independent detection of \siii, \mgii, \cii, and \feii\ absorption in FIRE and X-shooter data, and we simultaneously fit the \civ, \oi, and \siii\ systems across three instruments. Although the proximity of the \civ\ doublet in wavelength space introduces an inherent degeneracy with the \oi\ and \siii\ pair, high-resolution HIRES data reveal sufficient kinematic substructure to warrant an additional component at a mean velocity offset of $v_{\mathrm{offset}} = -85.4~\mathrm{kms^{-1}}$. We interpret this component as \oi, supported by the aforementioned coincident detections. To constrain the system, we use the secondary lines to fix the velocity centroid of the \oi\ component and perform simultaneous Voigt profile fitting across all three instruments, including both \civ\ transitions and single-component \oi\ and \siii\ models. The column density reported in this work corresponds to this singular \oi\ component, and the system is marked with an asterisk in Table~\ref{tab:logN_table}.

\subsection{\oi\ Absorption Detection Completeness}
\label{sec:completeness}

We quantify the completeness of \oi\ absorption detection around each galaxy as follows. For each galaxy overdensity, we compute the detection sensitivity on a pixel-by-pixel basis following the general framework of e.g.; \cite{Lanzetta_1987,Bernet_2010,mathes_2017}. Specifically, we use a continuum-normalized spectral slice of $\pm 1500$ \kms\ centered on the expected \oi\ absorption at the redshift of each galaxy overdensity.

We define the completeness in \oi\ absorption detection in a single instrument as

\begin{equation} C^{1}_{\text{OD}}(z) = H(W_{\min} - N_{\sigma} \frac{\sigma_{\text{EW}}(z)}{(1+z)}) \label{eqn:completeness} \end{equation}

where $H$ is the Heaviside step function and $W_{\min}$ is the equivalent-width cutoff at which completeness is defined. We adopt $W_{\min} = 0.05$\AA\ to facilitate comparison with the literature (see Section~\ref{sec:dNdX}). $N_{\sigma}$ is the detection threshold, which we set to $N_{\sigma} = 3$ by convention. The equivalent-width error, $\sigma_{\text{EW}}(z)$, is given by

\begin{equation} \sigma_{EW}(z)^{2} = \sum_{j}^{npix} \Delta\lambda_{j}^{2} (\sigma_{I_{j}}/I_{j}^{c})^{2}. \label{eqn:completeness_sigma} \end{equation}

where $\Delta\lambda_{j}$ is the wavelength spacing between adjacent pixels (set by the spectrograph), and $n_{\text{pix}}$ is the number of pixels over which the equivalent width is computed. We adopt the typical width of an \oi\ absorber as $n_{\text{pix}}$ for our analysis, yielding $n_{\text{pix, X-shooter}}=10$, $n_{\text{pix, HIRES}}=50$, and $n_{\text{pix, FIRE}}=11$, respectively. This is a conservative approach to calculating completeness as the typical width of the absorbers are larger than the resolving power of the instruments. We neglect continuum-fitting errors, so $\sigma_{\text{EW}}(z)$ reflects only statistical uncertainties from the spectra. Any chip gaps or sky lines are masked. $\sigma_{I}$ and $I^{c}$ are the flux uncertainty and continuum levels of the spectrum.

For galaxy overdensities observed with multiple spectrographs, we define the combined completeness as requiring detection in at least one instrument across the relevant wavelength range. For example, with three or more instruments available,

\begin{equation}
    C_{\text{OD}}(z) = \bigvee_{i} C^{i}_{\text{OD}}(z),
\end{equation}

where the logical OR ($\lor$) runs over all instruments $i \in {\text{X-shooter, FIRE, HIRES}}$. We then compute the mean completeness for each overdensity, and obtain the overall mean completeness, $\bar{C}$, by averaging over all overdensities weighted by their wavelength coverage. This yields a mean completeness of $\bar{C} =$ 91.4\% for \oi\ absorption around galaxy overdensities in the EIGER survey for $W_{\min} \geq $0.05 \AA.

\section{Results} \label{sec:results}
In this section we present the variation of \oi\ absorption around $z\sim6$ galaxies, incidence of \oi\ absorption per co-moving path length that is due to \oi\ absorption around galaxies and the chemical abundance of these systems. 

\subsection{Variation of \oi\ Absorption Around Galaxies}
\label{sec:variation_OI_around_galaxies}

Since all 40 galaxies in our sample reside within 15 identified overdensities (Section~\ref{sec:method_overdensities}), we investigate whether \oi\ absorption is preferentially detected in these environments. For clarity, we show the primary galaxy in each overdensity (as defined in Section~\ref{sec:method_overdensities}) when presenting the radial distribution of \oi\ absorption. Out of the 15 overdensities, five show associated \oi\ absorption. 

\begin{figure*}[!htb]
\centering
\includegraphics[width=1.0\linewidth]{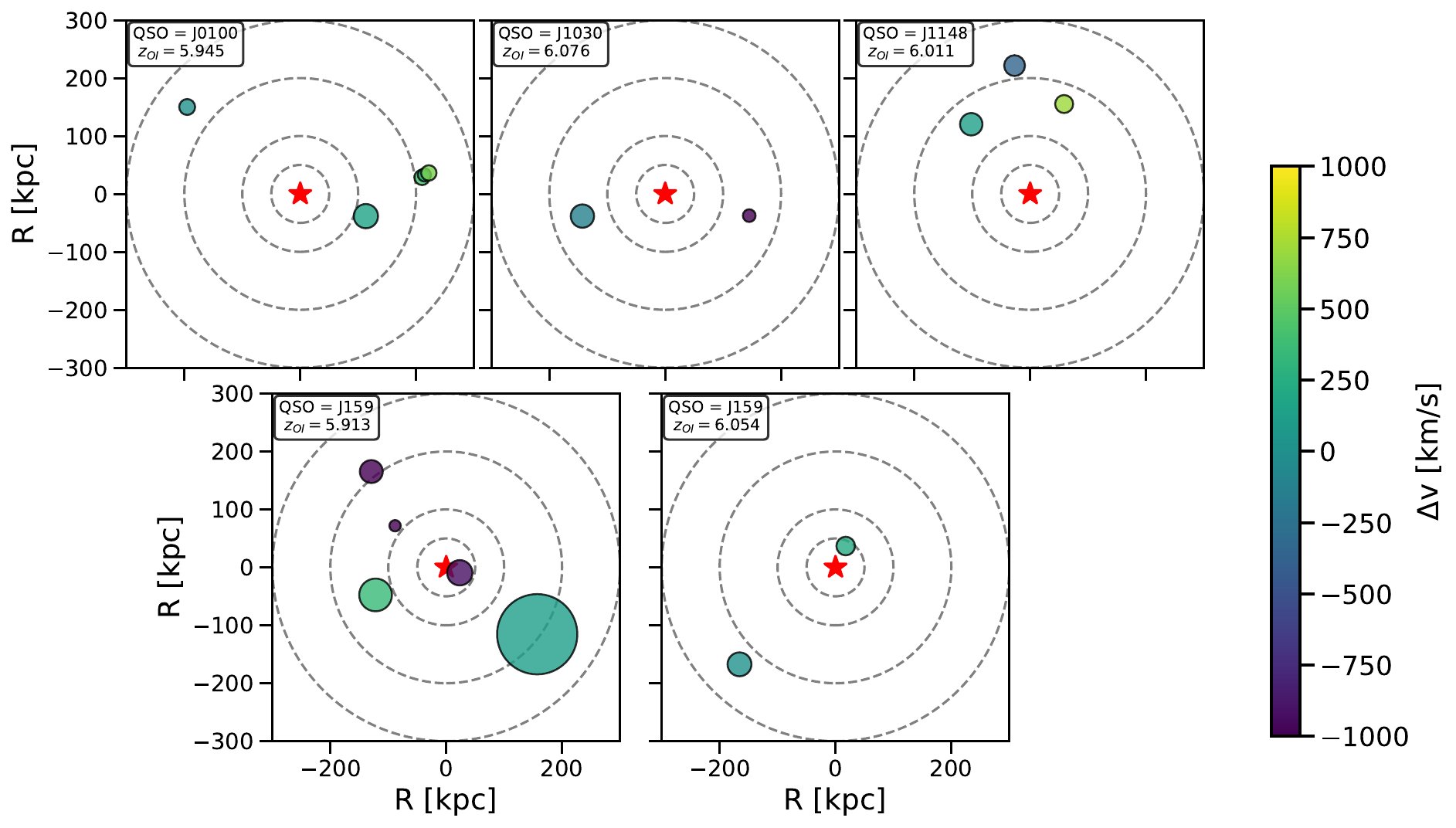}
\caption{\label{fig:OI_skymap} 
Galaxy overdensities around $z \sim 6$ \oi\ absorbers in the EIGER survey. Each panel shows the background quasar (red star) at the center, with foreground galaxies associated with \oi\ absorption marked as circles. Circles are color-coded by the velocity offset between the galaxy and the \oi\ absorber, and circle sizes are scaled to their virial radii. Only galaxies within 300 kpc of the quasar are shown, highlighting the overdensities nearest the absorbers. Dashed circles denote projected impact parameters of 50, 100, 200, and 300 pkpc from the quasar.}
\end{figure*}

Figure~\ref{fig:OI_skymap} shows the 2D spatial distribution of galaxies around these five detected systems, with the size depicting the virial radii of the galaxies. The galaxies are randomly distributed around the absorbers, and we do not find correlations with stellar mass, star formation, or other galaxy properties. In all five cases where we have \oi\ detection the galaxy member with the minimum impact parameter is also the member with the lowest $R/R_{vir}$.

\begin{figure*}[t] 
    \centering
    \includegraphics[width=0.95\textwidth]{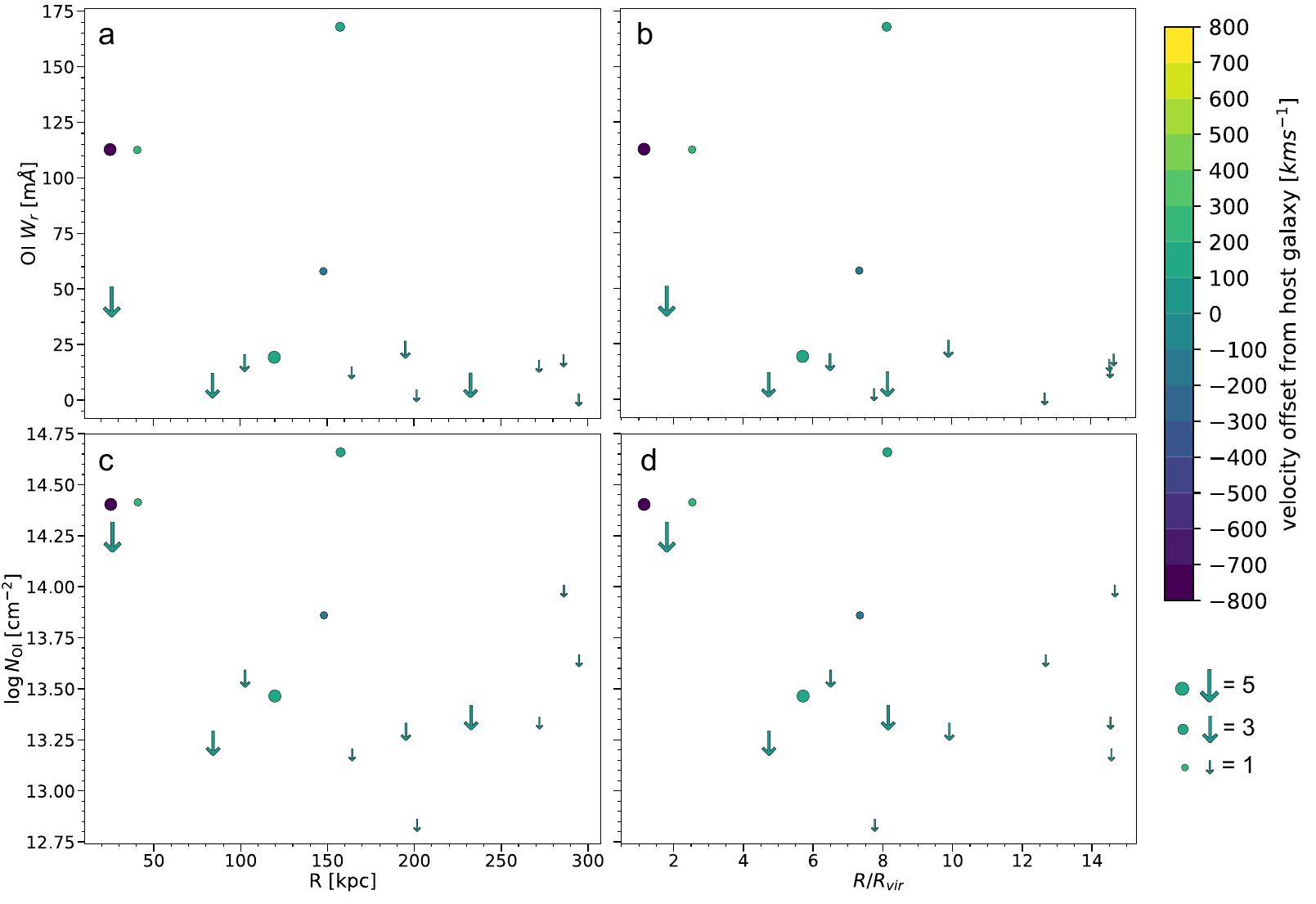}
    \caption{\label{fig:radial_profile} 
Radial absorption profiles of \oi\ absorption at $z \sim 6$. Each point marks the impact parameter to the nearest galaxy in a galaxy overdensity, with all detected \oi\ absorption arising in such environments. Overdensities with \oi\ absorption are shown as circles, while non-detections (2$\sigma$ limits) are indicated by downward arrows. Points are color-coded by the velocity offset between the galaxy redshift and the \oi\ absorber, and their size scales with the number of members in the overdensity. The mean uncertainty on the virial radius is $\Delta R_{\mathrm{vir}} = 1.81~\mathrm{kpc}$. \textit{Top panels:} Rest-frame equivalent width versus impact parameter ($R$, left) and normalized impact parameter ($R/R_{\mathrm{vir}}$, right). \textit{Bottom panels:} Same as above, but for \oi\ column densities.
}
\end{figure*}

Figure~\ref{fig:radial_profile} presents {\color{red}1D} the radial profiles around the primary galaxies in these systems. The marker size indicates the number of member galaxies in each overdensity, while the color encodes the velocity offset from the corresponding \oi\ absorption. Panels (a) and (c) show impact parameter ($R$), and panels (b) and (d) show the same distribution normalized by the halo virial radius ($R/R_{\rm vir}$).

Most of the \oi\ detections occur within $R \lesssim 175$ kpc of the primary galaxy, corresponding to an incidence rate of $0.56_{-0.16}^{+0.15}$, estimated using 68\% Wilson Score confidence intervals. Within 100 kpc, the incidence is $0.50_{-0.22}^{+0.22}$, increasing to $0.67_{-0.28}^{+0.20}$ for $R \leq 50$ kpc. Conversely, at $R > 150$ kpc an incidence of $0.13_{-0.08}^{+0.16}$ is measured. Within 300 kpc of the primary galaxy, the overall incidence is $0.27_{-0.10}^{+0.13}$, with values of $0.17_{-0.08}^{+0.13}$ for $50 \leq R \leq 300$ kpc and $0.18_{-0.09}^{+0.14}$ for $100 \leq R \leq 300$ kpc. These rates are substantially weaker than the incidence of strong \mgii\ absorption at $z \sim 6$ \citep{bordoloi_eigerIV_2024}, indicating that neutral oxygen preferentially traces a different phase of the circumgalactic environment.

\begin{deluxetable*}{cccccccccccc}
\tablecaption{\oi\ absorption lines and associated galaxy properties. \label{tab:logN_table}}
\tabletypesize{\footnotesize}
\setlength{\tabcolsep}{3pt}
\tablewidth{\linewidth}
\tablecolumns{11}
\tablehead{
\colhead{label} &
\colhead{Quasar} &
\colhead{$z_{\mathrm{gal}}$\tablenotemark{a}} &
\colhead{$R$} &
\colhead{$\langle v \rangle$\tablenotemark{b}} &
\colhead{\#\tablenotemark{c}} &
\colhead{$z_{\mathrm{O\,I}}$} &
\colhead{$\log N_{\mathrm{O\,I}}$} &
\colhead{$\log N_{\mathrm{Si\,II}}$} &
\colhead{$\log N_{\mathrm{C\,II}}$} &
\colhead{$\log N_{\mathrm{C\,IV}}$} &
\colhead{$\log N_{\mathrm{Si\,IV}}$} \\
\colhead{} &
\colhead{} &
\colhead{} &
\colhead{[kpc]} &
\colhead{[km s$^{-1}$]} &
\colhead{galaxies} &
\colhead{} &
\colhead{$[{\rm cm}^{-2}]$} &
\colhead{$[{\rm cm}^{-2}]$} &
\colhead{$[{\rm cm}^{-2}]$} &
\colhead{$[{\rm cm}^{-2}]$} &
\colhead{$[{\rm cm}^{-2}]$}
}

\colnumbers
\startdata
a & J0100 & 5.941 & 119.6 & 157  & 5 & 5.945 & $13.46_{-0.15}^{+0.11}$ & $12.71_{-0.07}^{+0.08}$ & $13.55_{-0.02}^{+0.02}$ & $14.24_{-0.09}^{+0.09}$ & $12.98_{-0.04}^{+0.04}$ \\
b & J1148 & 6.008 & 157.5 & 130  & 3 & 6.011 & $14.66_{-0.02}^{+0.02}$ & $13.48_{-0.05}^{+0.04}$\tablenotemark{*} & $14.35_{-0.02}^{+0.02}$ & $14.31_{-0.02}^{+0.02}$ & $13.25_{-0.06}^{+0.06}$ \\
c & J159  & 5.934 & 25.2  & -920 & 5 & 5.913 & $14.40_{-0.03}^{+0.03}$ & $13.77_{-0.09}^{+0.13}$ & $14.37_{-0.02}^{+0.02}$ & -- & -- \\
d & J159  & 6.049 & 40.8  & 230  & 2 & 6.054 & $14.41_{-0.03}^{+0.04}$ & $< 13.12$ & $14.29_{-0.04}^{+0.04}$ & -- & -- \\
e & J1030 & 6.075 & 147.9 & -131 & 2 & 6.076 & $13.86_{-0.08}^{+0.08}$\tablenotemark{**} & $13.33_{-0.09}^{+0.08}$ & $13.25_{-0.08}^{+0.08}$ & -- & -- \\
\enddata
\tablenotetext{a}{Redshift of the primary member of the overdensity.}
\tablenotetext{b}{Velocity offset of \oi\ relative to the primary member galaxy of the overdensity.}
\tablenotetext{c}{Number of galaxy members in the overdensity.}
\tablenotetext{*}{For this system $Si\,\textsc{ii}\,\lambda1304$ was not present, so the value given here is derived from fitting $Si\,\textsc{ii}\,\lambda1526$.}
\tablenotetext{**}{This system is blended; see discussion in Section~\ref{sec:methods_line_detection}.}
\tablecomments{The galaxies listed are the primary member of groups with detections at the lowest impact parameter from the host quasar.}
\end{deluxetable*}

Panels (b) and (d) reveal that \oi\ absorbers are typically located at much larger $R/R_{\rm vir}$ than strong low-ionization tracers such as \mgii. This distinction suggests that \oi\ absorption at $z \sim 6$ is not confined to the CGM of an individual galaxy but instead reflects the larger-scale gaseous environment of galaxy overdensities — what might be described as “cosmic ecosystems” rather than isolated halos. 

Comparing the stellar mass distributions of overdensities with and without \oi\ absorption, we find no statistically significant differences. The mean stellar mass of galaxies in overdensities with absorption is $\log(\langle M_{\star} \rangle/M_{\odot}) = 8.36$, compared to $\log(\langle M_{\star} \rangle/M_{\odot}) = 8.50$ for those without.

We emphasize that at $z \sim 6$, the large $R/R_{vir}$ values of these halos imply that \oi\ absorption typically arises at scales well beyond the typical CGM [$R/R_{vir} \lesssim 1$] \citep{tumlinson2017} probed by low-ionization lines at lower redshift. Unlike the low-$z$ CGM, where metal absorbers are closely tied to the halo of a single massive galaxy, the \oi\ absorbers studied here appear to be linked to extended networks of low-mass, young galaxies embedded in shared overdensities. Over cosmic time, these small systems and their diffuse gaseous environments are expected to merge and evolve into more massive systems, eventually forming Milky Way–like galaxies by $z=0$. Thus, neutral \oi\ at reionization is best understood not as a tracer of isolated galactic halos, but as a probe of the extended environments that connect galaxies, gas, and structure growth in the early universe.

Because \oi\ absorption traces dense, neutral gas and requires only minimal ionization corrections, we can place a conservative lower limit on the total oxygen mass associated with the EIGER galaxies. These estimates are intended as simple, order-of-magnitude calculations to provide intuition about the amount of neutral oxygen relative to the observed galaxy properties. Following the formalism of \citet{Bordoloi_COS_2014}, the oxygen mass contained in an annulus bounded by $R_{\rm in}$ and $R_{\rm out}$ is

\begin{equation}
M_{\rm O}
= \frac{1}{f_{\rm ion}}\, f_{c}\, \pi \left(R_{\rm out}^2 - R_{\rm in}^2\right)
\,\langle N_{\rm O\,I}\rangle \, m_{\rm O}\,
\label{eq:MO_annulus}
\end{equation}

where $R_{\rm in}$ and $R_{\rm out}$ are the inner and outer radii of the annulus, $f_{c}$ is the covering fraction (the incidence of \oi\ detections in that bin), $\langle N_{OI} \rangle$ is the mean \oi\ column density in the annulus, and $m_{\rm O}=16\,m_{p}$ is the mass of an oxygen atom, with $m_{p}$ being the mass of a proton. For a conservative lower limit we adopt $f_{\rm ion}=1$, where $f_{\rm ion}$ is the ionization fraction, (i.e., we assume that all of the oxygen mass is in the neutral phase).  We justify this assumption as follows: \oi\ is tightly coupled to \hi\ through charge exchange, and in dense gas the required ionization correction is minor, even under strong radiation fields such as the Fermi Bubbles \citep{Bordoloi_2017, Ashley_2022}. Under a typical metagalactic UV background, it is generally less than 10\% and does not significantly alter the inferred oxygen abundance. In addition, oxygen does not deplete onto dust grains, unlike Si or C. For these reasons, we consider the approximation $N_{O,total} \approx N_{OI}$ to be valid for our analysis.

We compute masses in two radial bins $0$--$100\,$kpc and $100$--$300\,$kpc, respectively. The total oxygen mass within radius $R_{\rm max}$ is obtained by summing the masses of the constituent annuli. This yields a strict lower limit of $\gtrsim$ 2$\times 10^6 M_\odot$ on the total oxygen mass within $R_{\rm max}$ = 300 kpc.

We compare the inferred oxygen mass in the extended environment of our sample with the total oxygen content in the interstellar medium (ISM) of a typical $z\sim 6$ galaxy. In computing this quantity, we make the following assumptions. First, we assume that all galaxies in this work lie on the typical $z\sim 6$ mass–metallicity relation, derived by stacking [\oiii] emitter galaxies \citep{Matthee_2022}, and we adopt the gas-phase metallicity from this relation for each galaxy.

Because direct measurements of the total $H_2$ and \hi\ gas content are not available at these redshifts, we extrapolate the molecular gas fraction from \cite{Tacconi2018} to $z\sim 6$ and adopt a conservative \hi\ gas fraction based on local observations \citep{Huang2012}. We note that high-$z$ galaxies are likely more gas-rich than their local counterparts; therefore, the inferred ISM gas masses should be considered lower limits.

With these caveats in mind, we estimate the total ISM oxygen mass as

\begin{equation}
M_{\rm O, ISM} \gtrsim X_{\rm O} , M_{\rm gas, total},
\end{equation}

where we define $M_{\rm gas, total} = M_{\rm H_2} + M_{\rm HI}$. Here, $X_{\rm O}$ is the gas-phase oxygen abundance derived from the mass–metallicity relation, and $M_{\rm H_2}$ and $M_{\rm HI}$ are the molecular and atomic gas masses derived as described above. For our galaxy sample, the mean stellar mass is $\log(M_\star/M_\odot) = 8.5$, corresponding to an ISM oxygen mass of $M_{\rm O, ISM} \gtrsim 6.8 \times 10^5~M_\odot$. Remarkably, this indicates that the typical ISM oxygen content is comparable to the oxygen mass in the extended environment around these galaxies, implying that the circumgalactic or intergalactic medium at $z \sim 6$ contains a substantial reservoir of oxygen. While these estimates depend on several specific assumptions, the inferred ISM and CGM oxygen masses appear to be of similar magnitude.

At low redshift, both $L^\ast$ and sub-$L^\ast$ galaxies similarly exhibit substantial metal content in their circumgalactic medium (CGM), with the CGM containing at least as much metal as the ISM of their host galaxies \citep{Tumlinson2011,Bordoloi_COS_2014}. A key difference is that, whereas the low-$z$ CGM is typically gravitationally bound to its host galaxy, at high redshift much of the gas exists in the extended environment of galaxy overdensities rather than being bound to individual galaxies \citep[see also][]{bordoloi_eigerIV_2024}.

\subsection{Expected Incidence of \oi\ Absorbers}
\label{sec:dNdX}
In this section we report the expected redshift-path incidence of \oi\ absorbers around [\oiii] emitting galaxies detected in the EIGER survey. We define the number of \oi\ absorbers associated with each galaxy overdensity per unit absorption distance above a detection threshold ($W_{OI} \geq$ 0.05\AA\ ) as

\begin{equation}
\frac{dN}{dX} (W_{OI} \geq 0.05 \AA) = \left( \frac{c}{H_0} \right) f_c  n_{\text{OD}}  \pi R_{\text{max}}^2 \left( \frac{1}{\bar{C}} \right).
\label{equation:dn_dx}
\end{equation}

Here, $R_{\text{max}} = 300$ kpc is the survey radius adopted in this work, and $\bar{C}$ is the mean \oi\ absorber detection completeness (see Section~\ref{sec:completeness}; 91.4\% for $W_{OI} \geq$ 0.05 \AA ). $n_{\text{OD}}$ is defined as

\begin{equation}
n_{\text{OD}} = \frac{n_{\text{gal}}}{\langle \text{overdensity size} \rangle}.
\end{equation}

$n_{\text{gal}}$ is the number density of [\oiii] emitting galaxies per unit comoving volume, estimated from the [\oiii] luminosity function at $z \sim 6$. And $\langle \text{overdensity size} \rangle$ is simply the mean number of galaxies in each galaxy overdensity empirically measured to be $\langle \text{overdensity size} \rangle = 2.67$. $f_c$ is the \oi\ absorption covering fraction within 300 kpc for $W_{OI} \geq$ 0.05 \AA\ ($f_c = 0.27^{+0.13}_{-0.10}$) (Covering fraction within 300 kpc is independent of the definition of the primary member).

To estimate the number density of galaxies per unit comoving volume ($n_{\text{gal}}$), we integrate the $z\sim 6$ [\oiii] luminosity function derived from the EIGER survey, with a minimum luminosity of $L_{\text{min}} = 10^{42.2}~\mathrm{ergs^{-1}}$ \citep{Matthee_2022}:

\begin{equation}
n_{\text{gal}} (>L_{\text{min}}) = \int_{L_{\text{min}}}^{\infty} \Phi(L) dL,
\end{equation}

where $\Phi(L)$ is the [\oiii] luminosity function defined as

\begin{equation}
\Phi(L)\,dL = \frac{\Phi^*}{L^*}\left(\frac{L}{L^*}\right)^{\alpha}\exp\!\left(-\frac{L}{L^*}\right)\,dL.
\end{equation}

Here, $\Phi^*$ is the characteristic number density, $L^*$ is the characteristic luminosity, and $\alpha$ is the faint-end slope. We adopt $\Phi^* = 10^{-4.02}~\mathrm{Mpc^{-3}}$, $L^* = 10^{43.33}~\mathrm{ergs^{-1}}$, and $\alpha = -1.92$, values derived from detailed completeness calculations of the EIGER [\oiii]-emitting galaxies (Mackenzie et al., in prep). This results in $n_{\text{gal}} = 8.586\times 10^{-04}\; \text{Mpc}^{-3}$. Using these values in equation \ref{equation:dn_dx} yields  $dN/dX =\; 0.12^{+0.06}_{-0.04}$, for \oi\ absorption around galaxies at a mean galaxy redshift of $6.001$.

All calculations in this section are performed with a detection threshold of $W_{\text{OI}} > 0.05\AA$, since the survey is highly complete down to this equivalent-width limit and because most of the literature surveys we compare against adopt the same threshold.

\begin{figure}[]
\centering
\includegraphics[width=1.0\linewidth]{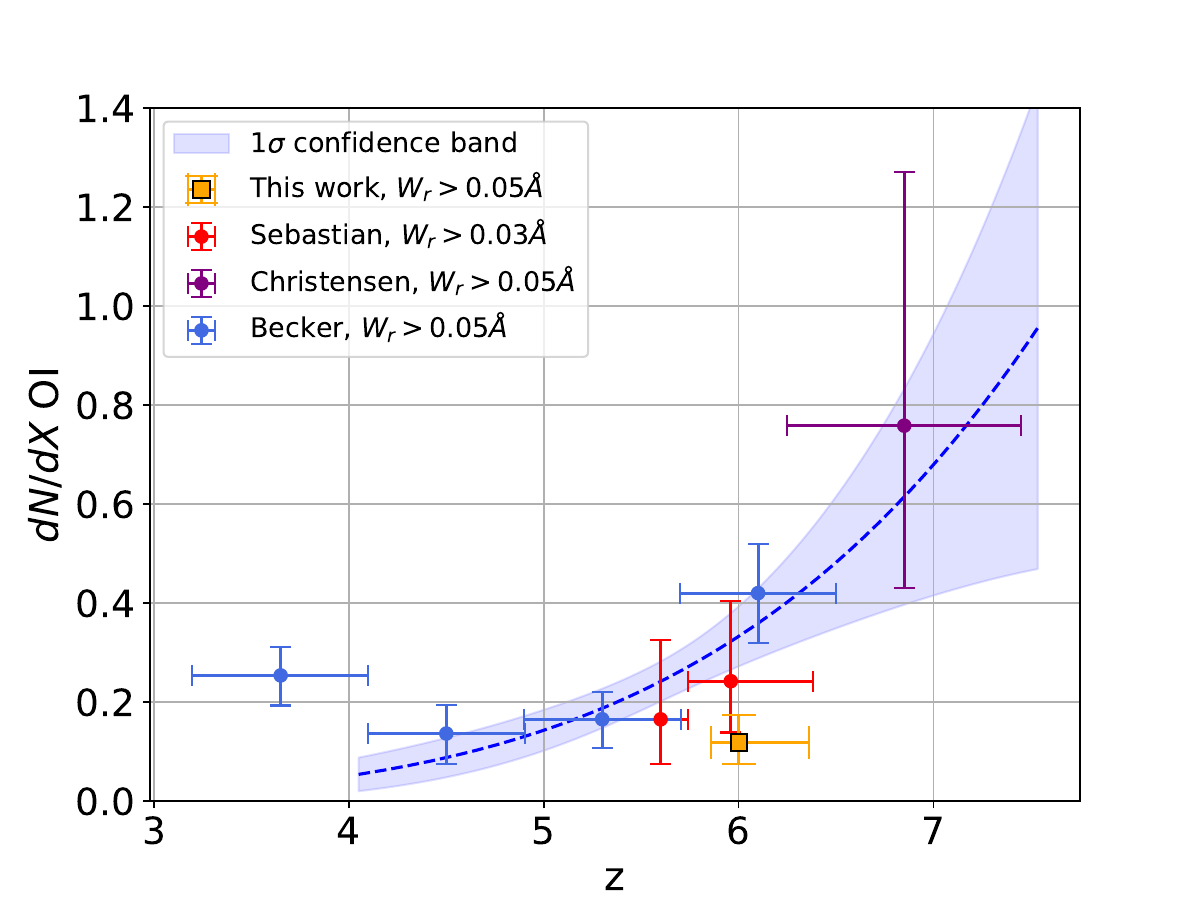}
\caption{\label{fig:dNdX_power_law} 
Incidence of \oi\ absorbers per comoving path length, $dN/dX$. Values are derived from all \oi\ absorption lines along high-$z$ quasar sightlines. Literature measurements are shown from \citet{Becker_2019} (blue circles), \citet{christensen_metal_2023} (purple circles), and \citet{sebastian2023} (red circles). The blue dashed line shows a power-law fit to the $z>4$ literature data, with the shaded region marking the $1\sigma$ uncertainty. The orange square gives the inferred $dN/dX$ when including only \oi\ absorption associated with host galaxies (see Section~\ref{sec:dNdX}), indicating that only $\sim35\%$ of \oi\ absorbers at $z\sim6$ arise in galaxy environments.
}

\end{figure}

We emphasize that this approach differs from blind QSO absorption-line surveys \citep[e.g.,][]{Becker_2006,sebastian2023,christensen_metal_2023,Sodini_2024}, where all \oi\ absorption-line systems along the full quasar redshift path are counted to compute $dN/dX$ directly, under the assumption that the absorber cross-section is contributed by all galaxies. In contrast, our method begins by defining the absorber cross-section around galaxy overdensities and then estimating the corresponding contribution to $dN/dX$ from the galaxy overdensities under study. Comparing these two approaches constrains the fraction of \oi\ (or other metal) absorption systems that arise in the extended halos of these galaxies. Figure~\ref{fig:dNdX_power_law} shows these literature measurements as colored circles, while the contribution to $dN/dX$ around $z\sim 6$ galaxy overdensities is indicated by an orange square. To quantify the relative contribution of galaxy overdensities, we fit a single power law to all literature values at $z>4$ (blue dashed line with shaded region). Comparing the $dN/dX$ inferred from quasar absorption-line surveys at $z\sim 6$ to our measured value, we find that \oi\ $dN/dX$ around $z\sim 6$ EIGER galaxy environments account for $35\% \pm 16\%$ of all \oi\ absorption-line systems detected in blind quasar surveys. This implies that [\oiii] emitting galaxy environments at $z\sim 6$ contribute roughly one-third of the \oi\ absorption systems at that epoch. Although the EIGER QSOs were selected to have known absorption systems along their sightlines, the EIGER galaxy survey around these sightlines is highly complete and independent of this selection. Therefore, we can create an unbiased measurement of the absorption cross-section ($dN/dX$) around these galaxy overdensities. We justify this comparison to blind absorption surveys in Section~\ref{sec:observations}.

This in turn suggests that $\sim 65\%$ of the \oi\ absorption systems detected along quasar sightlines at the epoch of reionization are not associated with [\oiii] emitter galaxy overdensities. Either the bulk of \oi\ absorption resides in isolated pockets of neutral IGM gas outside extended galaxy halos, or it is contributed by even lower-luminosity, less massive galaxies ($\log M*/M_{\odot} < 7.5$) that fall below the EIGER detection limits. Such systems may also be detected in galaxy halos that are oxygen-poor or lack strong [\oiii] emission features, since our galaxy identifications primarily rely on [\oiii] and \hbeta\ emission lines.

Simulations suggest that \oi\ absorption should primarily arise in galactic halos rather than in the diffuse IGM, since metals are not widely distributed and ionization fronts propagate more quickly than metals \citep{Oppenheimer_2009,Finlator_2013}. Very low-mass proto-galaxies forming their first stars may blur the boundary between IGM and CGM, enriching their surroundings with metals, especially if their star formation is bursty and prolific. If sufficiently numerous, such systems could contribute a non-negligible fraction of oxygen in the early Universe.  

Recent observations support this possibility. \citet{Nakajima_2025} reported a strongly lensed ($\mu=98$) galaxy at $z\approx6.6$ with a stellar mass upper limit of only $\sim 2700~M_\odot$ and a dynamical mass $\sim 100$ times larger. The source shows $[O_{III}]_{5008}$ emission and may be dominated by Population~III stars. This provides direct evidence of oxygen production in extremely low-mass systems. A population of such ultra-low mass dwarf galaxies could plausibly account for the $\sim$65\% of \oi\ absorbers detected along quasar sightlines that are not associated with galaxies in our survey. Our sensitivity limit of $\log M*/M_{\odot} \sim 7.5$ \citep{Matthee_2022} means such dwarfs remain undetected. However, these galaxies may lack a dense enough CGM to shield neutral \oi\ from the UV background; thus, even if they produce oxygen, neutral \oi\ may not survive \citep{Finlator_2013}. Interestingly, simulations by \cite{Kusmic_2024} showed that the contribution of metal ions, including \oi, is only weakly correlated with stellar mass of the galaxy, and the contribution is indistinguishable when comparing across galaxy luminosities. Additionally, they  argue that due the steep slope of the Stellar Mass Function (SMF), lower mass, lower luminosity galaxies contribute the majority of absorbing metal ions in the CGM and IGM.

Later simulation works by \cite{Doughty_2018, Doughty_2019} suggest that as the Epoch of Reionization continues, \oi\ will reduce overall but be preferentially found in higher column density regions close to halos, as they have a better ability to self-shield from ionizing photons. In fact, \cite{Doughty_2019} found that at $z=6$ the majority of \oi\ mass was located in the ISM of galaxies and only $\sim 30\%$ of \oi\ mass was found in the CGM and IGM. In contrast, in this work, we find that a substantial (comparable to ISM mass) Oxygen mass exists outside galaxies, however we stress that our mass measurements are order of magnitude estimates.
 A more direct comparison between these simulations and this work is the covering fraction. They report $\sim 5-25\%$ covering fraction at $z=6$ for \oi\ with $\log N > 14~\mathrm{cm}^{-2}$ out to $102~\mathrm{pkpc}$. In the EIGER survey, we report a covering fraction $f_c = 0.27^{+0.13}_{-0.10}$ using a minimum threshold of $W_{OI} \geq$ 0.05 \AA\ (or approximately $logN_{OI}>13.7~\mathrm{cm^{-2}}$) out to $300~\mathrm{kpc}$. Given the large measurement uncertainties and different definitions of associated galaxies, the covering fractions are qualitatively consistent between the two works.

The simulations of \citet{Finlator_2013} suggest that \oi\ absorbers with $\log N > 14~\mathrm{cm}^{-2}$ are most likely associated with halos of mass $\sim10^{9}~M_\odot$, where self-shielding is efficient. They further predict that only $\sim 13\%$ of all \oi\ absorption systems should reside in halos of $10^{10}$–$10^{11.5}~M_\odot$, the mass range of galaxies probed in this work. By contrast, we measure a higher $dN/dX$ of $\sim35\%$, but qualitatively our results remain consistent with the idea that roughly two-thirds of \oi\ absorbers are associated with undetected $\sim10^{9}~M_\odot$ halos. Future observations targeting lower-luminosity galaxies and higher redshifts will be essential to test these predictions.

After the initial posting of this manuscript, \citet{Pruto_2025} presented a simulation-based study that adopts the observational compilation shown in Figure~\ref{fig:dNdX_power_law} as a reference and explores the corresponding behavior in numerical models. Our minimum $W_{r}$ threshold corresponds to $\log N_{OI}>13.7~\mathrm{cm^{-2}}$; therefore, the most appropriate comparison is with their $\log N_{\mathrm{OI}} > 14~\mathrm{cm^{-2}}$ sample. Qualitatively, the two sets of results are in agreement. A key distinction lies in the definition of galactic environment. We characterize environments in terms of galaxy overdensities, which we argue is the most appropriate description of where \oi\ absorbers reside at $z \approx 6$, whereas \citet{Pruto_2025} define the CGM based on the virial radii of halos with $\log M > 8~M_{\odot}$. As a result, a direct quantitative comparison is not straightforward. Nevertheless, their inferred $dN/dX$ corresponds to an incidence rate of \oi\ absorbers of $\sim 58\% \pm 11\%$, which is consistent with our measurements within the uncertainties. This agreement may depend on their tighter constraints on absorber location and on the adopted minimum $W_r$ or $\log N$ threshold. \citet{Pruto_2025} further report that $\sim 80\%$ of \oi\ absorbers with $\log N_{\mathrm{OI}} > 13~\mathrm{cm^{-2}}$ are predicted to lie outside the CGM, also in qualitative agreement with our findings.

\begin{figure*}[]
\centering
\includegraphics[width=1.0\linewidth]{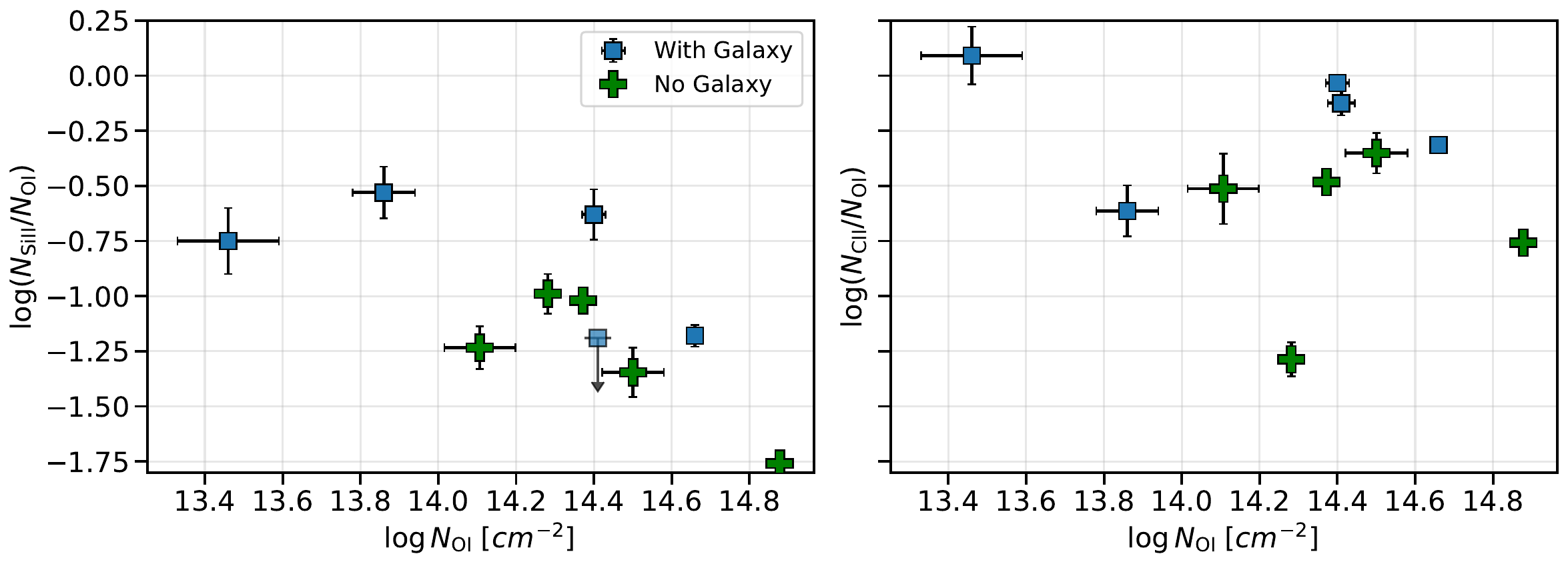}
\includegraphics[width=0.8\linewidth]{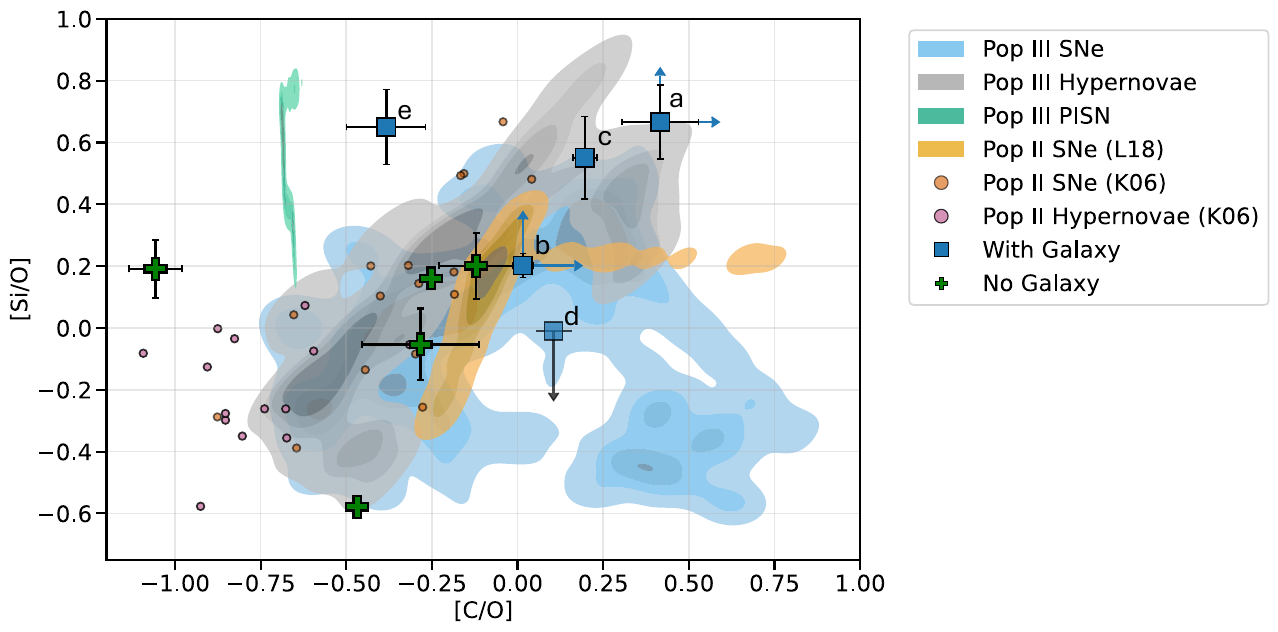}

\caption{\label{fig:SiII_OI_logN_ratio_figure} \textit{Top panels:} Ion column density ratios of \siii\ to \oi\ (left) and \cii\ to \oi\ (right) as a function of detected \oi\ column densities, shown in logarithmic units. Absorbers associated with galaxy overdensities (blue squares) and those not associated with galaxy overdensities (green crosses) exhibit clear variations in their column density ratios. One sightline with no detected \siii\ absorption is indicated with a downward arrow, where a $2\sigma$ upper limit on the column density is assumed. \textit{Bottom panel:} Relative abundance ratios [Si/O] versus [C/O] with respect to solar abundance. The background contours show the IMF-weighted abundance distributions from Population~II and Population~III supernova models. Orange and pink circles indicate abundance ratios predicted by low-metallicity Population~II supernova models. The abundance ratios of several \oi\ absorbers associated with galaxy overdensities are consistent with a significant contribution from Population~III supernovae.}
\end{figure*}

\subsection{Metal Abundances in \oi\ absorber population}

In this section we examine the relative abundances and metal column density ratios of \oi\ absorbers associated with galaxy overdensities, and compare them to \oi\ absorbers without any associated galaxies. To enable a fair comparison, we restrict our analysis to the six quasar sightlines from the EIGER survey, and identify all \oi\ absorption systems in the redshift range where an [\oiii]-emitting galaxy could be detected ($5.9 \leq z_{OI} \leq z_{QSO}$). We detect ten \oi\ absorption systems in total, five of which have associated galaxy overdensities (Table \ref{tab:logN_table}), and five without any identified galaxy association.

We first compare the ionic column density ratios in these two populations. Figure \ref{fig:SiII_OI_logN_ratio_figure} (top panels) shows the ratios of $N_{SiII}/N_{OI}$ and $N_{CII}/N_{OI}$ as functions of $N_{OI}$. Systems located near galaxy overdensities (blue squares) systematically show higher \siii/\oi\ and \cii/\oi\ ratios compared to systems without associated overdensities (green crosses). This offset could reflect differences in metallicity, or alternatively the effects of ionization. However, since the majority of these \oi\ systems lack detectable high-ionization transitions, it is reasonable to assume that neutral oxygen and singly ionized silicon and carbon dominate the elemental reservoirs \citep [see also][]{christensen_metal_2023}. In those cases, the observed ionic ratios provide direct tracers of relative abundances. Two exceptions are found along the J0100 and J1148 sightlines, which do exhibit \siiv\ and \civ\ absorption; for these systems the degeneracy between chemical composition and ionization corrections cannot be resolved. It is plausible that on average, \oi\ absorption found near galaxy overdensities have higher elemental abundance ratios as the non-galaxy associated absorbers either have significantly lower luminosity and less massive galaxies which would be expected to produce less metal, or there are truly no galaxies associated with this population.

We next compute the relative abundances [Si/O] and [C/O], presented in the bottom panel of Figure \ref{fig:SiII_OI_logN_ratio_figure}. These abundances are defined as
\begin{equation}
[\mathrm{Si/O}] = \log_{10} \left( \frac{N_{\mathrm{Si,tot}}}{N_{\mathrm{OI}}} \right) - \log_{10} \left( \frac{\mathrm{Si}}{\mathrm{O}} \right)_{\odot},
\end{equation}

\begin{equation}
[\mathrm{C/O}] = \log_{10} \left( \frac{N_{\mathrm{C,tot}}}{N_{\mathrm{OI}}} \right) - \log_{10} \left( \frac{\mathrm{C}}{\mathrm{O}} \right)_{\odot},
\end{equation}

where $N_{\mathrm{Si,tot}}$ and $N_{\mathrm{C,tot}}$ represent the summed ionic column densities. For systems with detected high-ionization absorption (\civ\ or \siiv), we include both low- and high-ionization states ($N_{\mathrm{SiII}}+N_{\mathrm{SiIV}}$ and $N_{\mathrm{CII}}+N_{\mathrm{CIV}}$), marking these measurements as lower limits (blue upward arrows in the figure) since additional undetected ionization states may be present. We note that in the two systems (a and b) exhibiting high-ionization absorption (see Table~\ref{tab:logN_table}), the low- and high-ionization absorption components are offset in velocity by $\Delta v$ =-66 \kms, and -69 \kms, respectively. These offsets are within $\lesssim$1.5 resolution elements of the instrument, and therefore we cannot statistically distinguish whether the low- and high-ionization components are kinematically distinct. This is suggestive of a complex, potentially multi-phase gaseous environment in the outskirts of these galaxies. For systems without detected high-ionization transitions, we adopt $N_{\mathrm{SiII}}$ and $N_{\mathrm{CII}}$ as the dominant contributions. We assume $N_{\mathrm{O,tot}} \approx N_{\mathrm{OI}}$, as ionization corrections for \oi\ are typically very small at high \oi\ column densities (see Section~\ref{sec:variation_OI_around_galaxies} for discussion). We adopt solar reference abundances of $\log_{10}(\mathrm{Si/O})_{\odot} = -1.18 \pm 0.05$ and $\log_{10}(\mathrm{C/O})_{\odot} = -0.23 \pm 0.06$ \citep{Asplund2021}.
 
Figure~\ref{fig:SiII_OI_logN_ratio_figure} (bottom panel) reveals a striking separation between \oi\ absorbers associated with galaxy overdensities (blue squares) and those without galaxy associations (green crosses). Systems near galaxies are found in the upper left of the [Si/O]–[C/O] diagram, with two cases exhibiting high-ionization absorption (lower abundance limits, blue arrows) which will offset them even farther from the locus of non-galaxy absorbers.

We place these abundance patterns in the context of chemical enrichment models from different stellar populations \citep[see also][]{christensen_metal_2023}. The contours and colored circles in Figure~\ref{fig:SiII_OI_logN_ratio_figure} highlight predicted [C/O] and [Si/O] yields from representative Population~II and Population~III supernova models, weighted by a Salpeter initial mass function (IMF). The weighting by number of stars predicted by an IMF is done to represent the typical observational relative abundance distribution in these models. These are not exhaustive but span the dominant theoretical expectations.

Population~III pair-instability SNe (PISNe; 140–260$M_{\odot}$; light green contours) predict narrow [C/O] $\simeq -0.7$ and high [Si/O] values, inconsistent with any of our observed systems. Such PISNe are also expected to produce substantial amounts of iron-peak elements, leading to elevated Fe/O ratios \citep{Tumlinson2004}. We will investigate \mgii\ and \feii\ absorption associated with $z\sim6$ galaxies in an upcoming work (Higginson et al., in prep.). Models of massive Population~III stars (10–100 $M_{\odot}$) produce both hypernovae (gray contours; $E>10^{52}$ erg) and standard core-collapse SNe (light blue contours) \citep{Heger2010}, spanning a broader region overlapping some of our measurements. We also consider Population~II models: low-metallicity core-collapse SNe from \citet{Limongi2018} (orange contours; Population~II SNe (L18)) occupy $-0.25 \lesssim [C/O] \lesssim 0.5$ and $-0.4 \lesssim [Si/O] \lesssim 0.4$, overlapping parts of the Population~III range, while the \citet{Kobayashi2006} yields (pink and orange circles for hypernovae and SNe, respectively) tend toward lower [Si/O] and [C/O].

Against these predictions, the contrast between the two absorber populations is clear. All non-galaxy \oi\ systems (green crosses) fall at the lower left of the diagram, consistent with enrichment from both  Population~II and/or Population~III SNe models. This regime also overlaps with ultra-metal-poor DLAs at $z\sim3$ ($-0.5 \lesssim [Si/O] \lesssim 0$, $-0.5 \lesssim [C/O] \lesssim 0$; \citealt{Cooke2011,Cooke2017}). By contrast, three of the five galaxy-associated absorbers lie well outside both the locus of Population~II predictions and the locus of ultra-metal-poor DLA relative abundances. These systems are best explained if Population~III enrichment played a major role, suggesting that they may represent a distinct population from ultra-metal-poor DLA host galaxies \citep[see also][]{Sodini_2024}.

Similarly, \cite{Sodini_2024} found that the [C/O] and [Si/O] abundance ratios for high-redshift \oi\ absorber systems are consistent with significant enrichment from Population~III, with the majority explained by $\geq30\%$ contributions. They also find a large spread in the implied Population~III contribution: several systems require $\geq60\%$ contributions, while others fall significantly below the average.  This is a strong indication of heterogeneous chemical enrichment by Population~III stars during the epoch of reionization.

The absorber toward J159 at $z=5.91$ shows [C/O] $=0.20$, [Si/O] $=0.55$ (labeled $c$), and no evidence of high-ionization absorption. These abundance ratios are best explained by contributions from Population~III SNe or hypernovae. The absorber toward J1030 at $z=6.07$ ([C/O] $=-0.38$, [Si/O] $=0.65$, labeled $e$) also show no high-ionization absorption. The relative abundances in these two predominantly neutral systems are therefore quite robust. Interestingly, the J1030 system is not reproduced by any single Population~II or Population~III model, suggesting that it may have experienced multiple enrichment episodes, producing unusually high [Si/O] coupled with low [C/O].

Two cases require special mention. The absorber toward J0100 at $z=5.94$ shows [C/O] $\geq0.42$ and [Si/O] $\geq0.67$ (labeled $a$), with strong \civ\ and \siiv\ absorption. Since these are lower limits, ionization corrections (for carbon and silicon) would push this system further upward, strengthening the case for Population~III contributions. This observation is best explained by contribution from Population~III hypernovae models. The J1148 absorber at $z=6.01$ shows [C/O] $\geq0.02$ and [Si/O] $\geq0.20$ (labeled $b$), again with high-ionization lines; correcting for ionization would shift it diagonally upward in the diagram. Taken together, four of the five galaxy-associated absorbers show relative abundances that cannot be explained by Population~II models alone, pointing instead to significant enrichment from Population~III supernovae or hypernovae.

Thus, most galaxy-associated \oi\ systems carry signatures of Population~III enrichment.
These results suggest that \oi\ systems tracing galaxy overdensities at $z\sim6$ carry chemical signatures of Population~III enrichment, likely from multiple supernovae events rather than a single explosion. The elevated [Si/O] and [C/O] ratios in these environments imply that [\oiii]-emitting galaxies at this epoch may retain significant Population~III contributions, and that metal-line absorbers around such overdensities provide a powerful probe of the first generations of stars.

\subsection{Comparison to the $z\approx0$ Universe}
At $z\sim6$, we cannot directly detect the local environments where \oi\ absorption systems arise. However, at $z\sim 0$, there are instances where the environments giving rise to \oi\ absorption are well understood. Here we discuss several such instances and compare them with $z\sim6$ observations. Such comparisons can provide useful insights regarding the nature of local environments within which \oi\ systems exist at high-$z$. We focus on \oi\ absorption associated with local blue compact dwarf (BCD) galaxies and the Magellanic Stream (MS), owing to their low metallicities and well-characterized local environments.

There is evidence that \oi\ is found preferentially in the \hii\ regions of local BCDs, with their \hi\ regions being pristine or showing very low levels of enrichment \citep{Kunth_1994, Pettini_1995, Thuan_1997}. In particular, \cite{Thuan_1997} reports that, to varying degrees, low-ionization metal lines (e.g., \oi\ and \siii) reside in the ionized hydrogen regions of BCDs rather than the neutral hydrogen regions. They report $\log N(\mathrm{SiII}/\mathrm{OI})~[\mathrm{cm^{-2}}]=0.574$. Compared to the $z\sim6$ measurements in panel b of Figure~\ref{fig:SiII_OI_logN_ratio_figure}, the \oi\ column density of the BCD is comparable, but the \siii-to-\oi\ ratio is inconsistent with the $z\sim6$ values. On the other hand, \cite{Lebouteiller_2004} finds that the \hii\ region of the BCD I~Zw~36 has significantly higher metallicity than its \hi\ region. However, inferred from velocity offsets, \oi\ in that system is associated with the \hi\ region. The reported \oi\ column density of that BCD ($\log N(\mathrm{OI})~[\mathrm{cm^{-2}}]=16.80_{-1.01}^{+1.85}$) is approximately two orders of magnitude higher than the $z\sim6$ sample. Given the available evidence, we conclude that the observed $z\sim6$ \oi\ absorption systems are not similar to local BCD systems and are unlikely to originate in \hii\ regions.

The Magellanic Stream is a filamentary trail of gas in the Local Group. Multiphase absorption-line observations from \cite{Fox_2013} and \cite{Richter_2013} support the argument that the MS was tidally stripped from the Magellanic Clouds approximately $2~\mathrm{Gyr}$ ago. Inferences from the abundance ratios of alpha elements in the MS are best explained by contributions from massive stars via Type~II supernovae when the MS was still part of the Magellanic Clouds, although the presence of significant dust depletion makes further inferences into enrichment history difficult. An additional study of the MS \citep{Kumari_2015} reports abundance ratios in a compact high-velocity cloud (CHVC) that is believed to be associated with the Magellanic Stream. They find $[\mathrm{Si/O}] = -0.04\pm0.25$ and $[\mathrm{C/O}] = -0.09\pm0.23$. When comparing these with the $z\sim6$ observations presented in panel~c of Figure~\ref{fig:SiII_OI_logN_ratio_figure}, we see that the CHVC abundance ratios occupy a different region of parameter space than the $z\approx6$ \oi\ absorbers within identified galactic overdensities showing Population~III enrichment signatures (points a, c, e). This discrepancy provides evidence that these $z\sim6$ \oi\ absorbers with Population~III enrichment signatures are unlikely to arise in environments similar to the MS. However, the ionic ratios of \oi\ absorption systems not associated with galaxy overdensities (green crosses) are consistent with this measurement. With the caveat that Magellanic Stream measurements do contain dust depletion effects, it is plausible that such \oi\ systems may trace IGM structures with densities comparable to those of the Magellanic Stream at high-$z$.

\section{Conclusion}

We have conducted a survey of neutral \oi\ absorption systems associated with galaxies during the Epoch of Reionization in the EIGER survey, using six high-redshift quasars as background sightlines. Our study focuses on the distribution of neutral oxygen around $[O_{III}]_{5008}$-emitting galaxies within 300~kpc and their connection to galaxy overdensities. The main results are as follows:
\begin{itemize}
    \item Five \oi\ absorbers are detected among 15 galaxy overdensities, spanning $z = 5.91$–$6.16$ with a mean redshift $\langle z \rangle = 6.001$.
    \item The \oi\ covering fraction within 300 kpc of the closest galaxy is $0.27^{+0.13}_{-0.10}$, with $0.67^{+0.20}_{-0.28}$ for $R \leq 50$ kpc and $0.17^{+0.13}_{-0.08}$ for $50 \leq R \leq 300$ kpc, where covering fraction is calculated using the galaxy in an overdensity with the lowest impact parameter.
    
  \item We find $\gtrsim 2 \times 10^{6}\,M_{\odot}$ of oxygen within 300 kpc, a mass comparable to the ISM oxygen content of $z \sim 6$ galaxies — assuming minimal ionization correction (see Section~\ref{sec:variation_OI_around_galaxies} for justification) — suggesting that the circumgalactic environment already holds metal reservoirs rivaling those inside galaxies.

    \item \oi\ absorbers are detected as far as $\sim 8\,R/R_{\rm vir}$ from galaxies, in sharp contrast to low-ionization tracers such as \mgii, which are typically confined within $\lesssim R/R_{\rm vir}$.

    \item The incidence of galaxy-associated \oi\ absorbers is $\frac{dN}{dX} = 0.12^{+0.06}_{-0.04}$. Comparison with blind quasar absorption-line surveys indicates that these systems represent only $\sim 35\%$ of all \oi\ absorbers at $z \sim 6$, implying that the majority likely arise in lower-mass galaxies below current detection limits or in dense neutral IGM pockets that survived reionization.

    \item Relative abundance and ionization ratios ([Si/O], [C/O]) show that galaxy-associated \oi\ systems are chemically enriched, with signatures in several systems consistent with contributions from Population~III stars. In particular, we identify four galaxy overdensities whose outskirts exhibit absorption-line abundance ratios consistent with contribution from Population~III nucleosynthesis.
    
\end{itemize}
Overall, these results indicate that neutral \oi\ at $z\sim6$ primarily traces extended overdensity environments rather than the CGM of individual halos. Moreover, the chemical abundance patterns of galaxy-associated \oi\ systems suggest contributions from Population~III stars, likely arising from multiple supernova events rather than a single explosion. The elevated [Si/O] and [C/O] ratios imply that [\oiii]-emitting galaxies at this epoch may retain significant Population~III enrichment. Consequently, metal-line absorbers around such overdensities offer a powerful probe of the first generations of stars, providing insight into early chemical enrichment and the role of primordial stellar populations in shaping galaxy evolution during the Epoch of Reionization.

\begin{acknowledgments}
This work is based on observations made with the NASA/
ESA/CSA James Webb Space Telescope. The data were
obtained from the Mikulski Archive for Space Telescopes at
the Space Telescope Science Institute, which is operated by the
Association of Universities for Research in Astronomy, Inc.,
under NASA contract NAS 5-03127 for JWST. These
observations are associated with program \#1243. This paper
includes data gathered with the 6.5 m Magellan Telescopes
located at Las Campanas Observatory, Chile. This work was supported by a NASA Keck PI Data Award, administered by the NASA Exoplanet Science Institute. Data presented herein were obtained at the W. M. Keck Observatory from telescope time allocated to the National Aeronautics and Space Administration (NASA) through the agency’s scientific partnership with the California Institute
of Technology and the University of California. The Observatory was made possible by the generous financial support of the W. M. Keck Foundation. The JWST data presented in this article were obtained from the Mikulski Archive for Space Telescopes (MAST) at the Space Telescope Science Institute. The specific observations analyzed can be accessed via \dataset[doi: 10.17909/y94q-h212]{https://doi.org/10.17909/y94q-h212}. This work was partially supported by the National Science Foundation under grant number: AST-2108931. Jorryt Matthee and Ivan Kramarenko acknowledge funding from the European Union (ERC, AGENTS,  101076224).\end{acknowledgments}

\facilities{JWST(NIRCam), Magellan: Baade (FIRE),VLT:
Kueyen (X-shooter), and Keck:I (HIRES, MOSFIRE)}

\software{astropy \citep{2013A&A...558A..33A,2018AJ....156..123A,2022ApJ...935..167A}, rbcodes \citep{rbcodes}, rbvfit \citep{rbvfit}. }

\bibliography{sample7}{}

\begin{thebibliography}{}
\expandafter\ifx\csname natexlab\endcsname\relax\def\natexlab#1{#1}\fi
\providecommand{\url}[1]{\href{#1}{#1}}
\providecommand{\dodoi}[1]{doi:~\href{http://doi.org/#1}{\nolinkurl{#1}}}
\providecommand{\doeprint}[1]{\href{http://ascl.net/#1}{\nolinkurl{http://ascl.net/#1}}}
\providecommand{\doarXiv}[1]{\href{https://arxiv.org/abs/#1}{\nolinkurl{https://arxiv.org/abs/#1}}}

\bibitem[{T. {Ashley} {et~al.}(2022){Ashley}, {Fox}, {Cashman}, {Lockman},
  {Bordoloi}, {Jenkins}, {Wakker}, \& {Karim}}]{Ashley_2022}
{Ashley}, T., {Fox}, A.~J., {Cashman}, F.~H., {et~al.} 2022,
  \bibinfo{title}{{Diverse metallicities of Fermi bubble clouds indicate dual
  origins in the disk and halo},} Nature Astronomy, 6, 968,
  \dodoi{10.1038/s41550-022-01720-0}

\bibitem[{M. {Asplund} {et~al.}(2021){Asplund}, {Amarsi}, \&
  {Grevesse}}]{Asplund2021}
{Asplund}, M., {Amarsi}, A.~M., \& {Grevesse}, N. 2021, \bibinfo{title}{{The
  chemical make-up of the Sun: A 2020 vision},} \aap, 653, A141,
  \dodoi{10.1051/0004-6361/202140445}

\bibitem[{ {Astropy Collaboration} {et~al.}(2013){Astropy Collaboration},
  {Robitaille}, {Tollerud}, {Greenfield}, {Droettboom}, {Bray}, {Aldcroft},
  {Davis}, {Ginsburg}, {Price-Whelan}, {Kerzendorf}, {Conley}, {Crighton},
  {Barbary}, {Muna}, {Ferguson}, {Grollier}, {Parikh}, {Nair}, {Unther},
  {Deil}, {Woillez}, {Conseil}, {Kramer}, {Turner}, {Singer}, {Fox}, {Weaver},
  {Zabalza}, {Edwards}, {Azalee Bostroem}, {Burke}, {Casey}, {Crawford},
  {Dencheva}, {Ely}, {Jenness}, {Labrie}, {Lim}, {Pierfederici}, {Pontzen},
  {Ptak}, {Refsdal}, {Servillat}, \& {Streicher}}]{2013A&A...558A..33A}
{Astropy Collaboration}, {Robitaille}, T.~P., {Tollerud}, E.~J., {et~al.} 2013,
  \bibinfo{title}{{Astropy: A community Python package for astronomy},} \aap,
  558, A33, \dodoi{10.1051/0004-6361/201322068}

\bibitem[{ {Astropy Collaboration} {et~al.}(2018){Astropy Collaboration},
  {Price-Whelan}, {Sip{\H{o}}cz}, {G{\"u}nther}, {Lim}, {Crawford}, {Conseil},
  {Shupe}, {Craig}, {Dencheva}, {Ginsburg}, {VanderPlas}, {Bradley},
  {P{\'e}rez-Su{\'a}rez}, {de Val-Borro}, {Aldcroft}, {Cruz}, {Robitaille},
  {Tollerud}, {Ardelean}, {Babej}, {Bach}, {Bachetti}, {Bakanov}, {Bamford},
  {Barentsen}, {Barmby}, {Baumbach}, {Berry}, {Biscani}, {Boquien}, {Bostroem},
  {Bouma}, {Brammer}, {Bray}, {Breytenbach}, {Buddelmeijer}, {Burke},
  {Calderone}, {Cano Rodr{\'\i}guez}, {Cara}, {Cardoso}, {Cheedella}, {Copin},
  {Corrales}, {Crichton}, {D'Avella}, {Deil}, {Depagne}, {Dietrich}, {Donath},
  {Droettboom}, {Earl}, {Erben}, {Fabbro}, {Ferreira}, {Finethy}, {Fox},
  {Garrison}, {Gibbons}, {Goldstein}, {Gommers}, {Greco}, {Greenfield},
  {Groener}, {Grollier}, {Hagen}, {Hirst}, {Homeier}, {Horton}, {Hosseinzadeh},
  {Hu}, {Hunkeler}, {Ivezi{\'c}}, {Jain}, {Jenness}, {Kanarek}, {Kendrew},
  {Kern}, {Kerzendorf}, {Khvalko}, {King}, {Kirkby}, {Kulkarni}, {Kumar},
  {Lee}, {Lenz}, {Littlefair}, {Ma}, {Macleod}, {Mastropietro}, {McCully},
  {Montagnac}, {Morris}, {Mueller}, {Mumford}, {Muna}, {Murphy}, {Nelson},
  {Nguyen}, {Ninan}, {N{\"o}the}, {Ogaz}, {Oh}, {Parejko}, {Parley}, {Pascual},
  {Patil}, {Patil}, {Plunkett}, {Prochaska}, {Rastogi}, {Reddy Janga},
  {Sabater}, {Sakurikar}, {Seifert}, {Sherbert}, {Sherwood-Taylor}, {Shih},
  {Sick}, {Silbiger}, {Singanamalla}, {Singer}, {Sladen}, {Sooley},
  {Sornarajah}, {Streicher}, {Teuben}, {Thomas}, {Tremblay}, {Turner},
  {Terr{\'o}n}, {van Kerkwijk}, {de la Vega}, {Watkins}, {Weaver}, {Whitmore},
  {Woillez}, {Zabalza}, \& {Astropy Contributors}}]{2018AJ....156..123A}
{Astropy Collaboration}, {Price-Whelan}, A.~M., {Sip{\H{o}}cz}, B.~M., {et~al.}
  2018, \bibinfo{title}{{The Astropy Project: Building an Open-science Project
  and Status of the v2.0 Core Package},} \aj, 156, 123,
  \dodoi{10.3847/1538-3881/aabc4f}

\bibitem[{ {Astropy Collaboration} {et~al.}(2022){Astropy Collaboration},
  {Price-Whelan}, {Lim}, {Earl}, {Starkman}, {Bradley}, {Shupe}, {Patil},
  {Corrales}, {Brasseur}, {N{\"o}the}, {Donath}, {Tollerud}, {Morris},
  {Ginsburg}, {Vaher}, {Weaver}, {Tocknell}, {Jamieson}, {van Kerkwijk},
  {Robitaille}, {Merry}, {Bachetti}, {G{\"u}nther}, {Aldcroft},
  {Alvarado-Montes}, {Archibald}, {B{\'o}di}, {Bapat}, {Barentsen},
  {Baz{\'a}n}, {Biswas}, {Boquien}, {Burke}, {Cara}, {Cara}, {Conroy},
  {Conseil}, {Craig}, {Cross}, {Cruz}, {D'Eugenio}, {Dencheva}, {Devillepoix},
  {Dietrich}, {Eigenbrot}, {Erben}, {Ferreira}, {Foreman-Mackey}, {Fox},
  {Freij}, {Garg}, {Geda}, {Glattly}, {Gondhalekar}, {Gordon}, {Grant},
  {Greenfield}, {Groener}, {Guest}, {Gurovich}, {Handberg}, {Hart},
  {Hatfield-Dodds}, {Homeier}, {Hosseinzadeh}, {Jenness}, {Jones}, {Joseph},
  {Kalmbach}, {Karamehmetoglu}, {Ka{\l}uszy{\'n}ski}, {Kelley}, {Kern},
  {Kerzendorf}, {Koch}, {Kulumani}, {Lee}, {Ly}, {Ma}, {MacBride}, {Maljaars},
  {Muna}, {Murphy}, {Norman}, {O'Steen}, {Oman}, {Pacifici}, {Pascual},
  {Pascual-Granado}, {Patil}, {Perren}, {Pickering}, {Rastogi}, {Roulston},
  {Ryan}, {Rykoff}, {Sabater}, {Sakurikar}, {Salgado}, {Sanghi}, {Saunders},
  {Savchenko}, {Schwardt}, {Seifert-Eckert}, {Shih}, {Jain}, {Shukla}, {Sick},
  {Simpson}, {Singanamalla}, {Singer}, {Singhal}, {Sinha}, {Sip{\H{o}}cz},
  {Spitler}, {Stansby}, {Streicher}, {{\v{S}}umak}, {Swinbank}, {Taranu},
  {Tewary}, {Tremblay}, {de Val-Borro}, {Van Kooten}, {Vasovi{\'c}}, {Verma},
  {de Miranda Cardoso}, {Williams}, {Wilson}, {Winkel}, {Wood-Vasey}, {Xue},
  {Yoachim}, {Zhang}, {Zonca}, \& {Astropy Project
  Contributors}}]{2022ApJ...935..167A}
{Astropy Collaboration}, {Price-Whelan}, A.~M., {Lim}, P.~L., {et~al.} 2022,
  \bibinfo{title}{{The Astropy Project: Sustaining and Growing a
  Community-oriented Open-source Project and the Latest Major Release (v5.0) of
  the Core Package},} \apj, 935, 167, \dodoi{10.3847/1538-4357/ac7c74}

\bibitem[{G.~D. Becker {et~al.}(2015)Becker, Bolton, \&
  Lidz}]{Becker_Bolton_Lidz_2015}
Becker, G.~D., Bolton, J.~S., \& Lidz, A. 2015, \bibinfo{title}{Reionisation
  and High-Redshift Galaxies: The View from Quasar Absorption Lines,}
  Publications of the Astronomical Society of Australia, 32, e045,
  \dodoi{10.1017/pasa.2015.45}

\bibitem[{G.~D. {Becker} {et~al.}(2006){Becker}, {Sargent}, {Rauch}, \&
  {Simcoe}}]{Becker_2006}
{Becker}, G.~D., {Sargent}, W. L.~W., {Rauch}, M., \& {Simcoe}, R.~A. 2006,
  \bibinfo{title}{{Discovery of Excess O I Absorption toward the z=6.42 QSO
  SDSS J1148+5251},} \apj, 640, 69, \dodoi{10.1086/500079}

\bibitem[{G.~D. {Becker} {et~al.}(2019){Becker}, {Pettini}, {Rafelski},
  {D'Odorico}, {Boera}, {Christensen}, {Cupani}, {Ellison}, {Farina},
  {Fumagalli}, {L{\'o}pez}, {Neeleman}, {Ryan-Weber}, \&
  {Worseck}}]{Becker_2019}
{Becker}, G.~D., {Pettini}, M., {Rafelski}, M., {et~al.} 2019,
  \bibinfo{title}{{The Evolution of O I over 3.2 < z < 6.5: Reionization of the
  Circumgalactic Medium},} \apj, 883, 163, \dodoi{10.3847/1538-4357/ab3eb5}

\bibitem[{P. {Behroozi} {et~al.}(2019){Behroozi}, {Wechsler}, {Hearin}, \&
  {Conroy}}]{Behroozi_2019}
{Behroozi}, P., {Wechsler}, R.~H., {Hearin}, A.~P., \& {Conroy}, C. 2019,
  \bibinfo{title}{{UNIVERSEMACHINE: The correlation between galaxy growth and
  dark matter halo assembly from z = 0-10},} \mnras, 488, 3143,
  \dodoi{10.1093/mnras/stz1182}

\bibitem[{M.~L. Bernet {et~al.}(2010)Bernet, Miniati, \& Lilly}]{Bernet_2010}
Bernet, M.~L., Miniati, F., \& Lilly, S.~J. 2010, \bibinfo{title}{Mg ii
  ABSORPTION SYSTEMS WITH W0 ⩾ 0.1 Å FOR A RADIO SELECTED SAMPLE OF 77
  QUASI-STELLAR OBJECTS AND THEIR ASSOCIATED MAGNETIC FIELDS AT HIGH
  REDSHIFT*,} The Astrophysical Journal, 711, 380,
  \dodoi{10.1088/0004-637X/711/1/380}

\bibitem[{R. Bordoloi \& J. Higginson(2025)Bordoloi \& Higginson}]{rbvfit}
Bordoloi, R., \& Higginson, J. 2025, \bibinfo{title}{rbvfit: A Voigt profile
  fitting tool,}, 2.0.0 Zenodo, \dodoi{10.5281/zenodo.16318060}

\bibitem[{R. Bordoloi {et~al.}(2025)Bordoloi, Liu, Clark, Higginson, \&
  Flores}]{rbcodes}
Bordoloi, R., Liu, B., Clark, S., Higginson, J., \& Flores, D. 2025,
  \bibinfo{title}{rongmon/rbcodes: rbcodes v1.0.0,}, v1.0.0 Zenodo,
  \dodoi{10.5281/zenodo.6079263}

\bibitem[{R. {Bordoloi} {et~al.}(2011){Bordoloi}, {Lilly}, {Knobel},
  {Bolzonella}, {Kampczyk}, {Carollo}, {Iovino}, {Zucca}, {Contini}, {Kneib},
  {Le Fevre}, {Mainieri}, {Renzini}, {Scodeggio}, {Zamorani}, {Balestra},
  {Bardelli}, {Bongiorno}, {Caputi}, {Cucciati}, {de la Torre}, {de Ravel},
  {Garilli}, {Kova{\v{c}}}, {Lamareille}, {Le Borgne}, {Le Brun}, {Maier},
  {Mignoli}, {Pello}, {Peng}, {Perez Montero}, {Presotto}, {Scarlata},
  {Silverman}, {Tanaka}, {Tasca}, {Tresse}, {Vergani}, {Barnes}, {Cappi},
  {Cimatti}, {Coppa}, {Diener}, {Franzetti}, {Koekemoer}, {L{\'o}pez-Sanjuan},
  {McCracken}, {Moresco}, {Nair}, {Oesch}, {Pozzetti}, \&
  {Welikala}}]{Bordoloi2011}
{Bordoloi}, R., {Lilly}, S.~J., {Knobel}, C., {et~al.} 2011,
  \bibinfo{title}{{The Radial and Azimuthal Profiles of Mg II Absorption around
  0.5 < z < 0.9 zCOSMOS Galaxies of Different Colors, Masses, and
  Environments},} \apj, 743, 10, \dodoi{10.1088/0004-637X/743/1/10}

\bibitem[{R. {Bordoloi} {et~al.}(2014){Bordoloi}, {Tumlinson}, {Werk},
  {Oppenheimer}, {Peeples}, {Prochaska}, {Tripp}, {Katz}, {Dav{\'e}}, {Fox},
  {Thom}, {Ford}, {Weinberg}, {Burchett}, \& {Kollmeier}}]{Bordoloi_COS_2014}
{Bordoloi}, R., {Tumlinson}, J., {Werk}, J.~K., {et~al.} 2014,
  \bibinfo{title}{{The COS-Dwarfs Survey: The Carbon Reservoir around Sub-L*
  Galaxies},} \apj, 796, 136, \dodoi{10.1088/0004-637X/796/2/136}

\bibitem[{R. {Bordoloi} {et~al.}(2017){Bordoloi}, {Fox}, {Lockman}, {Wakker},
  {Jenkins}, {Savage}, {Hernandez}, {Tumlinson}, {Bland-Hawthorn}, \&
  {Kim}}]{Bordoloi_2017}
{Bordoloi}, R., {Fox}, A.~J., {Lockman}, F.~J., {et~al.} 2017,
  \bibinfo{title}{{Mapping the Nuclear Outflow of the Milky Way: Studying the
  Kinematics and Spatial Extent of the Northern Fermi Bubble},} \apj, 834, 191,
  \dodoi{10.3847/1538-4357/834/2/191}

\bibitem[{R. {Bordoloi} {et~al.}(2024){Bordoloi}, {Simcoe}, {Matthee},
  {Kashino}, {Mackenzie}, {Lilly}, {Eilers}, {Liu}, {DePalma}, {Yue}, \& {P.
  Naidu}}]{bordoloi_eigerIV_2024}
{Bordoloi}, R., {Simcoe}, R.~A., {Matthee}, J., {et~al.} 2024,
  \bibinfo{title}{{EIGER IV. The Cool {}10$^{4}$ K Circumgalactic Environment
  of High-redshift Galaxies Reveals Remarkably Efficient Intergalactic Medium
  Enrichment},} \apj, 963, 28, \dodoi{10.3847/1538-4357/ad1b63}

\bibitem[{S.~E.~I. {Bosman} {et~al.}(2022){Bosman}, {Davies}, {Becker},
  {Keating}, {Davies}, {Zhu}, {Eilers}, {D'Odorico}, {Bian}, {Bischetti},
  {Cristiani}, {Fan}, {Farina}, {Haehnelt}, {Hennawi}, {Kulkarni}, {Mesinger},
  {Meyer}, {Onoue}, {Pallottini}, {Qin}, {Ryan-Weber}, {Schindler}, {Walter},
  {Wang}, \& {Yang}}]{Bosman_2022}
{Bosman}, S. E.~I., {Davies}, F.~B., {Becker}, G.~D., {et~al.} 2022,
  \bibinfo{title}{{Hydrogen reionization ends by z = 5.3:
  Lyman-{\ensuremath{\alpha}} optical depth measured by the XQR-30 sample},}
  \mnras, 514, 55, \dodoi{10.1093/mnras/stac1046}

\bibitem[{M. {Brooks} {et~al.}(2025){Brooks}, {Simons}, {Trump}, {Taylor},
  {Bagley}, {Backhaus}, {Davis}, {Buat}, {Cleri}, {de la Vega}, {Finkelstein},
  {Hirschmann}, {Holwerda}, {Kocevski}, {Koekemoer}, {Lucas}, {Pacucci}, \&
  {Seill{\'e}}}]{Brooks_2025}
{Brooks}, M., {Simons}, R.~C., {Trump}, J.~R., {et~al.} 2025,
  \bibinfo{title}{{Here There Be (Dusty) Monsters: High-redshift Active
  Galactic Nuclei Are Dustier than Their Hosts},} \apj, 986, 177,
  \dodoi{10.3847/1538-4357/addac4}

\bibitem[{J.~N. {Burchett} {et~al.}(2019){Burchett}, {Tripp}, {Prochaska},
  {Werk}, {Tumlinson}, {Howk}, {Willmer}, {Lehner}, {Meiring}, {Bowen},
  {Bordoloi}, {Peeples}, {Jenkins}, {O'Meara}, {Tejos}, \&
  {Katz}}]{Burchett_2019}
{Burchett}, J.~N., {Tripp}, T.~M., {Prochaska}, J.~X., {et~al.} 2019,
  \bibinfo{title}{{The COS Absorption Survey of Baryon Harbors (CASBaH):
  Warm-Hot Circumgalactic Gas Reservoirs Traced by Ne VIII Absorption},} \apjl,
  877, L20, \dodoi{10.3847/2041-8213/ab1f7f}

\bibitem[{D. {Calzetti} {et~al.}(2000){Calzetti}, {Armus}, {Bohlin}, {Kinney},
  {Koornneef}, \& {Storchi-Bergmann}}]{Calzetti2000}
{Calzetti}, D., {Armus}, L., {Bohlin}, R.~C., {et~al.} 2000,
  \bibinfo{title}{{The Dust Content and Opacity of Actively Star-forming
  Galaxies},} \apj, 533, 682, \dodoi{10.1086/308692}

\bibitem[{G. {Chabrier}(2003){Chabrier}}]{Chabrier2003}
{Chabrier}, G. 2003, \bibinfo{title}{{Galactic Stellar and Substellar Initial
  Mass Function},} \pasp, 115, 763, \dodoi{10.1086/376392}

\bibitem[{J. {Choi} {et~al.}(2016){Choi}, {Dotter}, {Conroy}, {Cantiello},
  {Paxton}, \& {Johnson}}]{Choi2016}
{Choi}, J., {Dotter}, A., {Conroy}, C., {et~al.} 2016, \bibinfo{title}{{Mesa
  Isochrones and Stellar Tracks (MIST). I. Solar-scaled Models},} \apj, 823,
  102, \dodoi{10.3847/0004-637X/823/2/102}

\bibitem[{L. {Christensen} {et~al.}(2023){Christensen}, {Jakobsen}, {Willott},
  {Arribas}, {Bunker}, {Charlot}, {Maiolino}, {Marshall}, {Perna}, \&
  {{\"U}bler}}]{christensen_metal_2023}
{Christensen}, L., {Jakobsen}, P., {Willott}, C., {et~al.} 2023,
  \bibinfo{title}{{Metal enrichment and evolution in four z > 6.5 quasar
  sightlines observed with JWST/NIRSpec},} \aap, 680, A82,
  \dodoi{10.1051/0004-6361/202347943}

\bibitem[{R. {Cooke} {et~al.}(2011){Cooke}, {Pettini}, {Steidel}, {Rudie}, \&
  {Nissen}}]{Cooke2011}
{Cooke}, R., {Pettini}, M., {Steidel}, C.~C., {Rudie}, G.~C., \& {Nissen},
  P.~E. 2011, \bibinfo{title}{{The most metal-poor damped
  Ly{\ensuremath{\alpha}} systems: insights into chemical evolution in the very
  metal-poor regime},} \mnras, 417, 1534,
  \dodoi{10.1111/j.1365-2966.2011.19365.x}

\bibitem[{R.~J. {Cooke} {et~al.}(2017){Cooke}, {Pettini}, \&
  {Steidel}}]{Cooke2017}
{Cooke}, R.~J., {Pettini}, M., \& {Steidel}, C.~C. 2017,
  \bibinfo{title}{{Discovery of the most metal-poor damped
  Lyman-{\ensuremath{\alpha}} system},} \mnras, 467, 802,
  \dodoi{10.1093/mnras/stx037}

\bibitem[{T.~J. {Cooper} {et~al.}(2019){Cooper}, {Simcoe}, {Cooksey},
  {Bordoloi}, {Miller}, {Furesz}, {Turner}, \& {Ba{\~n}ados}}]{Cooper_2019}
{Cooper}, T.~J., {Simcoe}, R.~A., {Cooksey}, K.~L., {et~al.} 2019,
  \bibinfo{title}{{Heavy Element Absorption Systems at 5.0 < z < 6.8:
  Metal-poor Neutral Gas and a Diminishing Signature of Highly Ionized
  Circumgalactic Matter},} \apj, 882, 77, \dodoi{10.3847/1538-4357/ab3402}

\bibitem[{A. {D'Aloisio} {et~al.}(2017){D'Aloisio}, {Upton Sanderbeck},
  {McQuinn}, {Trac}, \& {Shapiro}}]{Daloisio_2017}
{D'Aloisio}, A., {Upton Sanderbeck}, P.~R., {McQuinn}, M., {Trac}, H., \&
  {Shapiro}, P.~R. 2017, \bibinfo{title}{{On the contribution of active
  galactic nuclei to the high-redshift metagalactic ionizing background},}
  \mnras, 468, 4691, \dodoi{10.1093/mnras/stx711}

\bibitem[{R.~L. {Davies} {et~al.}(2023{\natexlab{a}}){Davies}, {Ryan-Weber},
  {D'Odorico}, {Bosman}, {Meyer}, {Becker}, {Cupani}, {Bischetti}, {Sebastian},
  {Eilers}, {Farina}, {Wang}, {Yang}, \& {Zhu}}]{Davies_2023a}
{Davies}, R.~L., {Ryan-Weber}, E., {D'Odorico}, V., {et~al.}
  2023{\natexlab{a}}, \bibinfo{title}{{The XQR-30 metal absorber catalogue: 778
  absorption systems spanning 2 {\ensuremath{\lesssim}} z
  {\ensuremath{\lesssim}} 6.5},} \mnras, 521, 289,
  \dodoi{10.1093/mnras/stac3662}

\bibitem[{R.~L. {Davies} {et~al.}(2023{\natexlab{b}}){Davies}, {Ryan-Weber},
  {D'Odorico}, {Bosman}, {Meyer}, {Becker}, {Cupani}, {Keating}, {Bischetti},
  {Davies}, {Eilers}, {Farina}, {Haehnelt}, {Pallottini}, \&
  {Zhu}}]{Davies_2023b}
{Davies}, R.~L., {Ryan-Weber}, E., {D'Odorico}, V., {et~al.}
  2023{\natexlab{b}}, \bibinfo{title}{{Examining the decline in the C IV
  content of the Universe over 4.3 {\ensuremath{\lesssim}} z
  {\ensuremath{\lesssim}} 6.3 using the E-XQR-30 sample},} \mnras, 521, 314,
  \dodoi{10.1093/mnras/stad294}

\bibitem[{C.~G. {D{\'\i}az} {et~al.}(2021){D{\'\i}az}, {Ryan-Weber}, {Karman},
  {Caputi}, {Salvadori}, {Crighton}, {Ouchi}, \& {Vanzella}}]{Diaz_2021}
{D{\'\i}az}, C.~G., {Ryan-Weber}, E.~V., {Karman}, W., {et~al.} 2021,
  \bibinfo{title}{{Faint LAEs near z > 4.7 C IV absorbers revealed by MUSE},}
  \mnras, 502, 2645, \dodoi{10.1093/mnras/staa3129}

\bibitem[{V. {D'Odorico} {et~al.}(2022){D'Odorico}, {Finlator}, {Cristiani},
  {Cupani}, {Perrotta}, {Calura}, {C{\`e}nturion}, {Becker}, {Berg}, {Lopez},
  {Ellison}, \& {Pomante}}]{Dordorico_2022}
{D'Odorico}, V., {Finlator}, K., {Cristiani}, S., {et~al.} 2022,
  \bibinfo{title}{{The evolution of the Si IV content in the Universe from the
  epoch of reionization to cosmic noon},} \mnras, 512, 2389,
  \dodoi{10.1093/mnras/stac545}

\bibitem[{A. {Dotter}(2016){Dotter}}]{Dotter2016}
{Dotter}, A. 2016, \bibinfo{title}{{MESA Isochrones and Stellar Tracks (MIST)
  0: Methods for the Construction of Stellar Isochrones},} \apjs, 222, 8,
  \dodoi{10.3847/0067-0049/222/1/8}

\bibitem[{C. {Doughty} \& K. {Finlator}(2019){Doughty} \&
  {Finlator}}]{Doughty_2019}
{Doughty}, C., \& {Finlator}, K. 2019, \bibinfo{title}{{Evolution of neutral
  oxygen during the epoch of reionization and its use in estimating the neutral
  hydrogen fraction},} \mnras, 489, 2755, \dodoi{10.1093/mnras/stz2331}

\bibitem[{C. {Doughty} {et~al.}(2018){Doughty}, {Finlator}, {Oppenheimer},
  {Dav{\'e}}, \& {Zackrisson}}]{Doughty_2018}
{Doughty}, C., {Finlator}, K., {Oppenheimer}, B.~D., {Dav{\'e}}, R., \&
  {Zackrisson}, E. 2018, \bibinfo{title}{{Aligned metal absorbers and the
  ultraviolet background at the end of reionization},} \mnras, 475, 4717,
  \dodoi{10.1093/mnras/sty156}

\bibitem[{A.-C. {Eilers} {et~al.}(2023){Eilers}, {Simcoe}, {Yue}, {Mackenzie},
  {Matthee}, {{\v{D}}urov{\v{c}}{\'\i}kov{\'a}}, {Kashino}, {Bordoloi}, \&
  {Lilly}}]{Eilers2023}
{Eilers}, A.-C., {Simcoe}, R.~A., {Yue}, M., {et~al.} 2023,
  \bibinfo{title}{{EIGER. III. JWST/NIRCam Observations of the Ultraluminous
  High-redshift Quasar J0100+2802},} \apj, 950, 68,
  \dodoi{10.3847/1538-4357/acd776}

\bibitem[{C.-A.
  {Faucher-Gigu{\`e}re}(2020){Faucher-Gigu{\`e}re}}]{Giguere_2020}
{Faucher-Gigu{\`e}re}, C.-A. 2020, \bibinfo{title}{{A cosmic UV/X-ray
  background model update},} \mnras, 493, 1614, \dodoi{10.1093/mnras/staa302}

\bibitem[{C.-A. {Faucher-Gigu{\`e}re} \& S.~P. {Oh}(2023){Faucher-Gigu{\`e}re}
  \& {Oh}}]{Faucher_2023}
{Faucher-Gigu{\`e}re}, C.-A., \& {Oh}, S.~P. 2023, \bibinfo{title}{{Key
  Physical Processes in the Circumgalactic Medium},} \araa, 61, 131,
  \dodoi{10.1146/annurev-astro-052920-125203}

\bibitem[{K. {Finlator} {et~al.}(2013){Finlator}, {Mu{\~n}oz}, {Oppenheimer},
  {Oh}, {{\"O}zel}, \& {Dav{\'e}}}]{Finlator_2013}
{Finlator}, K., {Mu{\~n}oz}, J.~A., {Oppenheimer}, B.~D., {et~al.} 2013,
  \bibinfo{title}{{The host haloes of O I absorbers in the reionization
  epoch},} \mnras, 436, 1818, \dodoi{10.1093/mnras/stt1697}

\bibitem[{D. Foreman-Mackey {et~al.}(2013)Foreman-Mackey, Hogg, Lang, \&
  Goodman}]{emcee}
Foreman-Mackey, D., Hogg, D.~W., Lang, D., \& Goodman, J. 2013,
  \bibinfo{title}{emcee: The MCMC Hammer,} Publications of the Astronomical
  Society of the Pacific, 125, 306, \dodoi{10.1086/670067}

\bibitem[{A.~J. {Fox} {et~al.}(2013){Fox}, {Richter}, {Wakker}, {Lehner},
  {Howk}, {Ben Bekhti}, {Bland-Hawthorn}, \& {Lucas}}]{Fox_2013}
{Fox}, A.~J., {Richter}, P., {Wakker}, B.~P., {et~al.} 2013,
  \bibinfo{title}{{The COS/UVES Absorption Survey of the Magellanic Stream. I.
  One-tenth Solar Abundances along the Body of the Stream},} \apj, 772, 110,
  \dodoi{10.1088/0004-637X/772/2/110}

\bibitem[{S.~R. {Furlanetto} \& S.~P. {Oh}(2005){Furlanetto} \&
  {Oh}}]{furlanetto_2005}
{Furlanetto}, S.~R., \& {Oh}, S.~P. 2005, \bibinfo{title}{{Taxing the rich:
  recombinations and bubble growth during reionization},} \mnras, 363, 1031,
  \dodoi{10.1111/j.1365-2966.2005.09505.x}

\bibitem[{A. {Hamanowicz} {et~al.}(2020){Hamanowicz}, {P{\'e}roux}, {Zwaan},
  {Rahmani}, {Pettini}, {York}, {Klitsch}, {Augustin}, {Krogager}, {Kulkarni},
  {Fresco}, {Biggs}, {Milliard}, \& {Vernet}}]{Hamanowicz_2020}
{Hamanowicz}, A., {P{\'e}roux}, C., {Zwaan}, M.~A., {et~al.} 2020,
  \bibinfo{title}{{MUSE-ALMA haloes V: physical properties and environment of z
  {\ensuremath{\leq}} 1.4 H I quasar absorbers},} \mnras, 492, 2347,
  \dodoi{10.1093/mnras/stz3590}

\bibitem[{A. {Heger} \& S.~E. {Woosley}(2010){Heger} \& {Woosley}}]{Heger2010}
{Heger}, A., \& {Woosley}, S.~E. 2010, \bibinfo{title}{{Nucleosynthesis and
  Evolution of Massive Metal-free Stars},} \apj, 724, 341,
  \dodoi{10.1088/0004-637X/724/1/341}

\bibitem[{S. {Huang} {et~al.}(2012){Huang}, {Haynes}, {Giovanelli}, \&
  {Brinchmann}}]{Huang2012}
{Huang}, S., {Haynes}, M.~P., {Giovanelli}, R., \& {Brinchmann}, J. 2012,
  \bibinfo{title}{{The Arecibo Legacy Fast ALFA Survey: The Galaxy Population
  Detected by ALFALFA},} \apj, 756, 113, \dodoi{10.1088/0004-637X/756/2/113}

\bibitem[{L. Jiang {et~al.}(2022)Jiang, Ning, Fan, Ho, Luo, Wang, Wu, Wu, Yang,
  \& Zheng}]{Jiang_2022}
Jiang, L., Ning, Y., Fan, X., {et~al.} 2022, \bibinfo{title}{Definitive upper
  bound on the negligible contribution of quasars to cosmic reionization,}
  Nature Astronomy, 6, 850–856, \dodoi{10.1038/s41550-022-01708-w}

\bibitem[{B.~D. {Johnson} {et~al.}(2021){Johnson}, {Leja}, {Conroy}, \&
  {Speagle}}]{Johnson2021}
{Johnson}, B.~D., {Leja}, J., {Conroy}, C., \& {Speagle}, J.~S. 2021,
  \bibinfo{title}{{Stellar Population Inference with Prospector},} \apjs, 254,
  22, \dodoi{10.3847/1538-4365/abef67}

\bibitem[{J.~L. {Johnson} \& A. {Aykutalp}(2019){Johnson} \&
  {Aykutalp}}]{Johnson_2019}
{Johnson}, J.~L., \& {Aykutalp}, A. 2019, \bibinfo{title}{{Extreme Primordial
  Star Formation Enabled by High-redshift Quasars},} \apj, 879, 18,
  \dodoi{10.3847/1538-4357/ab223e}

\bibitem[{M. Karamanis \& F. Beutler(2020)Karamanis \& Beutler}]{ZEUS2}
Karamanis, M., \& Beutler, F. 2020, \bibinfo{title}{Ensemble slice sampling:
  Parallel, black-box and gradient-free inference for correlated & multimodal
  distributions,} arXiv preprint arXiv: 2002.06212

\bibitem[{M. Karamanis {et~al.}(2021)Karamanis, Beutler, \& Peacock}]{ZEUS1}
Karamanis, M., Beutler, F., \& Peacock, J.~A. 2021, \bibinfo{title}{zeus: A
  Python implementation of Ensemble Slice Sampling for efficient Bayesian
  parameter inference,} arXiv preprint arXiv:2105.03468

\bibitem[{D. {Kashino} {et~al.}(2023){Kashino}, {Lilly}, {Matthee}, {Eilers},
  {Mackenzie}, {Bordoloi}, \& {Simcoe}}]{Kashino2023}
{Kashino}, D., {Lilly}, S.~J., {Matthee}, J., {et~al.} 2023,
  \bibinfo{title}{{EIGER. I. A Large Sample of [O III]-emitting Galaxies at 5.3
  < z < 6.9 and Direct Evidence for Local Reionization by Galaxies},} \apj,
  950, 66, \dodoi{10.3847/1538-4357/acc588}

\bibitem[{D. {Kashino} {et~al.}(2025){Kashino}, {Lilly}, {Matthee},
  {Mackenzie}, {Eilers}, {Bordoloi}, {Simcoe}, {Naidu}, {Yue}, \&
  {Liu}}]{EIGERVII}
{Kashino}, D., {Lilly}, S.~J., {Matthee}, J., {et~al.} 2025,
  \bibinfo{title}{{EIGER VII. The evolving relationship between galaxies and
  the intergalactic medium in the final stages of reionization},} arXiv
  e-prints, arXiv:2506.03121, \dodoi{10.48550/arXiv.2506.03121}

\bibitem[{R.~S. Klessen \& S.~C.~O. Glover(2023)Klessen \&
  Glover}]{klessen_2023_annual_review}
Klessen, R.~S., \& Glover, S. C.~O. 2023, \bibinfo{title}{The first stars:
  formation, properties, and impact,} \doarXiv{2303.12500}

\bibitem[{C. {Kobayashi} {et~al.}(2006){Kobayashi}, {Umeda}, {Nomoto},
  {Tominaga}, \& {Ohkubo}}]{Kobayashi2006}
{Kobayashi}, C., {Umeda}, H., {Nomoto}, K., {Tominaga}, N., \& {Ohkubo}, T.
  2006, \bibinfo{title}{{Galactic Chemical Evolution: Carbon through Zinc},}
  \apj, 653, 1145, \dodoi{10.1086/508914}

\bibitem[{G. {Kulkarni} {et~al.}(2019){Kulkarni}, {Worseck}, \&
  {Hennawi}}]{Kulkarni_2019}
{Kulkarni}, G., {Worseck}, G., \& {Hennawi}, J.~F. 2019,
  \bibinfo{title}{{Evolution of the AGN UV luminosity function from redshift
  7.5},} \mnras, 488, 1035, \dodoi{10.1093/mnras/stz1493}

\bibitem[{N. {Kumari} {et~al.}(2015){Kumari}, {Fox}, {Tumlinson}, {Thom},
  {Westmeier}, \& {Ely}}]{Kumari_2015}
{Kumari}, N., {Fox}, A.~J., {Tumlinson}, J., {et~al.} 2015, \bibinfo{title}{{A
  Compact High Velocity Cloud near the Magellanic Stream: Metallicity and
  Small-scale Structure},} \apj, 800, 44, \dodoi{10.1088/0004-637X/800/1/44}

\bibitem[{D. {Kunth} {et~al.}(1994){Kunth}, {Lequeux}, {Sargent}, \&
  {Viallefond}}]{Kunth_1994}
{Kunth}, D., {Lequeux}, J., {Sargent}, W.~L.~W., \& {Viallefond}, F. 1994,
  \bibinfo{title}{{Is there primordial gas in IZw 18 ?},} \aap, 282, 709

\bibitem[{S. {Ku{\v{s}}mi{\'c}} {et~al.}(2024){Ku{\v{s}}mi{\'c}}, {Finlator},
  {Huscher}, \& {Steen}}]{Kusmic_2024}
{Ku{\v{s}}mi{\'c}}, S., {Finlator}, K., {Huscher}, E., \& {Steen}, M. 2024,
  \bibinfo{title}{{Galaxy{\textendash}Absorber Association in the Epoch of
  Reionization: Galactic Population Luminosity Distribution for Different
  Absorbers at 10 {\ensuremath{\geq}} z {\ensuremath{\geq}} 5.5},} \apj, 974,
  224, \dodoi{10.3847/1538-4357/ad713c}

\bibitem[{K.~M. {Lanzetta} {et~al.}(1987){Lanzetta}, {Turnshek}, \&
  {Wolfe}}]{Lanzetta_1987}
{Lanzetta}, K.~M., {Turnshek}, D.~A., \& {Wolfe}, A.~M. 1987,
  \bibinfo{title}{{An Absorption-Line Survey of 32 QSOs at Red Wavelengths:
  Properties of the MG II Absorbers},} \apj, 322, 739, \dodoi{10.1086/165769}

\bibitem[{V. {Lebouteiller} {et~al.}(2004){Lebouteiller}, {Kunth}, {Lequeux},
  {Lecavelier des Etangs}, {D{\'e}sert}, {H{\'e}brard}, \&
  {Vidal-Madjar}}]{Lebouteiller_2004}
{Lebouteiller}, V., {Kunth}, D., {Lequeux}, J., {et~al.} 2004,
  \bibinfo{title}{{Abundance differences between the neutral and the ionized
  gas of the dwarf galaxy IZw 36},} \aap, 415, 55,
  \dodoi{10.1051/0004-6361:20034592}

\bibitem[{M. {Limongi} \& A. {Chieffi}(2018){Limongi} \&
  {Chieffi}}]{Limongi2018}
{Limongi}, M., \& {Chieffi}, A. 2018, \bibinfo{title}{{Presupernova Evolution
  and Explosive Nucleosynthesis of Rotating Massive Stars in the Metallicity
  Range -3 {\ensuremath{\leq}} [Fe/H] {\ensuremath{\leq}} 0},} \apjs, 237, 13,
  \dodoi{10.3847/1538-4365/aacb24}

\bibitem[{N.~L. {Mathes} {et~al.}(2017){Mathes}, {Churchill}, \&
  {Murphy}}]{mathes_2017}
{Mathes}, N.~L., {Churchill}, C.~W., \& {Murphy}, M.~T. 2017,
  \bibinfo{title}{{The Vulture Survey I: Analyzing the Evolution of ${\MgII}$
  Absorbers},} arXiv e-prints, arXiv:1701.05624,
  \dodoi{10.48550/arXiv.1701.05624}

\bibitem[{J. {Matthee} {et~al.}(2023){Matthee}, {Mackenzie}, {Simcoe},
  {Kashino}, {Lilly}, {Bordoloi}, \& {Eilers}}]{EIGERII}
{Matthee}, J., {Mackenzie}, R., {Simcoe}, R.~A., {et~al.} 2023,
  \bibinfo{title}{{EIGER. II. First Spectroscopic Characterization of the Young
  Stars and Ionized Gas Associated with Strong H{\ensuremath{\beta}} and [O
  III] Line Emission in Galaxies at z = 5-7 with JWST},} \apj, 950, 67,
  \dodoi{10.3847/1538-4357/acc846}

\bibitem[{J. Matthee {et~al.}(2022)Matthee, Naidu, Pezzulli, Gronke, Sobral,
  Oesch, Hayes, Erb, Schaerer, Amorín, Tacchella, Paulino-Afonso, Llerena,
  Calhau, \& Röttgering}]{Matthee_2022}
Matthee, J., Naidu, R.~P., Pezzulli, G., {et~al.} 2022,
  \bibinfo{title}{(Re)Solving reionization with Lyα: how bright Lyα Emitters
  account for the z ≈ 2–8 cosmic ionizing background,} Monthly Notices of
  the Royal Astronomical Society, 512, 5960–5977,
  \dodoi{10.1093/mnras/stac801}

\bibitem[{J. {Matthee} {et~al.}(2024){Matthee}, {Naidu}, {Brammer}, {Chisholm},
  {Eilers}, {Goulding}, {Greene}, {Kashino}, {Labbe}, {Lilly}, {Mackenzie},
  {Oesch}, {Weibel}, {Wuyts}, {Xiao}, {Bordoloi}, {Bouwens}, {van Dokkum},
  {Illingworth}, {Kramarenko}, {Maseda}, {Mason}, {Meyer}, {Nelson}, {Reddy},
  {Shivaei}, {Simcoe}, \& {Yue}}]{Matthee2024LRD}
{Matthee}, J., {Naidu}, R.~P., {Brammer}, G., {et~al.} 2024,
  \bibinfo{title}{{Little Red Dots: An Abundant Population of Faint Active
  Galactic Nuclei at z {\ensuremath{\sim}} 5 Revealed by the EIGER and FRESCO
  JWST Surveys},} \apj, 963, 129, \dodoi{10.3847/1538-4357/ad2345}

\bibitem[{R.~H. {Mebane} {et~al.}(2018){Mebane}, {Mirocha}, \&
  {Furlanetto}}]{Mebane_2018}
{Mebane}, R.~H., {Mirocha}, J., \& {Furlanetto}, S.~R. 2018,
  \bibinfo{title}{{The Persistence of Population III Star Formation},} \mnras,
  479, 4544, \dodoi{10.1093/mnras/sty1833}

\bibitem[{R.~P. {Naidu} {et~al.}(2024){Naidu}, {Matthee}, {Kramarenko},
  {Weibel}, {Brammer}, {Oesch}, {Lechner}, {Furtak}, {Di Cesare}, {Torralba},
  {Kotiwale}, {Bezanson}, {Bouwens}, {Chandra}, {Claeyssens}, {Danhaive},
  {Frebel}, {de Graaff}, {Greene}, {Heintz}, {Ji}, {Kashino}, {Katz}, {Labbe},
  {Leja}, {Li}, {Maseda}, {Richard}, {Shivaei}, {Simcoe}, {Sobral}, {Suess},
  {Tacchella}, \& {Williams}}]{Naidu_SF_model}
{Naidu}, R.~P., {Matthee}, J., {Kramarenko}, I., {et~al.} 2024,
  \bibinfo{title}{{All the Little Things in Abell 2744: $>$1000 Gravitationally
  Lensed Dwarf Galaxies at $z=0-9$ from JWST NIRCam Grism Spectroscopy},} arXiv
  e-prints, arXiv:2410.01874, \dodoi{10.48550/arXiv.2410.01874}

\bibitem[{K. {Nakajima} \& R. {Maiolino}(2022){Nakajima} \&
  {Maiolino}}]{Nakajima_2022}
{Nakajima}, K., \& {Maiolino}, R. 2022, \bibinfo{title}{{Diagnostics for PopIII
  galaxies and direct collapse black holes in the early universe},} \mnras,
  513, 5134, \dodoi{10.1093/mnras/stac1242}

\bibitem[{K. {Nakajima} {et~al.}(2025){Nakajima}, {Ouchi}, {Harikane},
  {Vanzella}, {Ono}, {Isobe}, {Nishigaki}, {Tsujimoto}, {Nakamura}, {Xu},
  {Umeda}, \& {Zhang}}]{Nakajima_2025}
{Nakajima}, K., {Ouchi}, M., {Harikane}, Y., {et~al.} 2025, \bibinfo{title}{{An
  Ultra-Faint, Chemically Primitive Galaxy Forming at the Epoch of
  Reionization},} arXiv e-prints, arXiv:2506.11846,
  \dodoi{10.48550/arXiv.2506.11846}

\bibitem[{B.~D. {Oppenheimer} {et~al.}(2009){Oppenheimer}, {Dav{\'e}}, \&
  {Finlator}}]{Oppenheimer_2009}
{Oppenheimer}, B.~D., {Dav{\'e}}, R., \& {Finlator}, K. 2009,
  \bibinfo{title}{{Tracing the re-ionization-epoch intergalactic medium with
  metal absorption lines},} \mnras, 396, 729,
  \dodoi{10.1111/j.1365-2966.2009.14771.x}

\bibitem[{Y. {Peng} {et~al.}(2015){Peng}, {Maiolino}, \&
  {Cochrane}}]{Peng_2015}
{Peng}, Y., {Maiolino}, R., \& {Cochrane}, R. 2015,
  \bibinfo{title}{{Strangulation as the primary mechanism for shutting down
  star formation in galaxies},} \nat, 521, 192, \dodoi{10.1038/nature14439}

\bibitem[{C. {Peroux} \& D. {Nelson}(2024){Peroux} \&
  {Nelson}}]{Peroux_Nelson_2024}
{Peroux}, C., \& {Nelson}, D. 2024, \bibinfo{title}{{The Multi-Scale
  Multi-Phase Circumgalactic Medium: Observed and Simulated},} arXiv e-prints,
  arXiv:2411.07988, \dodoi{10.48550/arXiv.2411.07988}

\bibitem[{M. {Pettini} \& K. {Lipman}(1995){Pettini} \&
  {Lipman}}]{Pettini_1995}
{Pettini}, M., \& {Lipman}, K. 1995, \bibinfo{title}{{On the oxygen abundance
  of neutral gas in I ZW 18.},} \aap, 297, L63,
  \dodoi{10.48550/arXiv.astro-ph/9503075}

\bibitem[{ {Planck Collaboration} {et~al.}(2020){Planck Collaboration},
  {Aghanim}, {Akrami}, {Ashdown}, {Aumont}, {Baccigalupi}, {Ballardini},
  {Banday}, {Barreiro}, {Bartolo}, {Basak}, {Battye}, {Benabed}, {Bernard},
  {Bersanelli}, {Bielewicz}, {Bock}, {Bond}, {Borrill}, {Bouchet}, {Boulanger},
  {Bucher}, {Burigana}, {Butler}, {Calabrese}, {Cardoso}, {Carron},
  {Challinor}, {Chiang}, {Chluba}, {Colombo}, {Combet}, {Contreras}, {Crill},
  {Cuttaia}, {de Bernardis}, {de Zotti}, {Delabrouille}, {Delouis}, {Di
  Valentino}, {Diego}, {Dor{\'e}}, {Douspis}, {Ducout}, {Dupac}, {Dusini},
  {Efstathiou}, {Elsner}, {En{\ss}lin}, {Eriksen}, {Fantaye}, {Farhang},
  {Fergusson}, {Fernandez-Cobos}, {Finelli}, {Forastieri}, {Frailis},
  {Fraisse}, {Franceschi}, {Frolov}, {Galeotta}, {Galli}, {Ganga},
  {G{\'e}nova-Santos}, {Gerbino}, {Ghosh}, {Gonz{\'a}lez-Nuevo}, {G{\'o}rski},
  {Gratton}, {Gruppuso}, {Gudmundsson}, {Hamann}, {Handley}, {Hansen},
  {Herranz}, {Hildebrandt}, {Hivon}, {Huang}, {Jaffe}, {Jones}, {Karakci},
  {Keih{\"a}nen}, {Keskitalo}, {Kiiveri}, {Kim}, {Kisner}, {Knox},
  {Krachmalnicoff}, {Kunz}, {Kurki-Suonio}, {Lagache}, {Lamarre}, {Lasenby},
  {Lattanzi}, {Lawrence}, {Le Jeune}, {Lemos}, {Lesgourgues}, {Levrier},
  {Lewis}, {Liguori}, {Lilje}, {Lilley}, {Lindholm}, {L{\'o}pez-Caniego},
  {Lubin}, {Ma}, {Mac{\'\i}as-P{\'e}rez}, {Maggio}, {Maino}, {Mandolesi},
  {Mangilli}, {Marcos-Caballero}, {Maris}, {Martin}, {Martinelli},
  {Mart{\'\i}nez-Gonz{\'a}lez}, {Matarrese}, {Mauri}, {McEwen}, {Meinhold},
  {Melchiorri}, {Mennella}, {Migliaccio}, {Millea}, {Mitra},
  {Miville-Desch{\^e}nes}, {Molinari}, {Montier}, {Morgante}, {Moss}, {Natoli},
  {N{\o}rgaard-Nielsen}, {Pagano}, {Paoletti}, {Partridge}, {Patanchon},
  {Peiris}, {Perrotta}, {Pettorino}, {Piacentini}, {Polastri}, {Polenta},
  {Puget}, {Rachen}, {Reinecke}, {Remazeilles}, {Renzi}, {Rocha}, {Rosset},
  {Roudier}, {Rubi{\~n}o-Mart{\'\i}n}, {Ruiz-Granados}, {Salvati}, {Sandri},
  {Savelainen}, {Scott}, {Shellard}, {Sirignano}, {Sirri}, {Spencer},
  {Sunyaev}, {Suur-Uski}, {Tauber}, {Tavagnacco}, {Tenti}, {Toffolatti},
  {Tomasi}, {Trombetti}, {Valenziano}, {Valiviita}, {Van Tent}, {Vibert},
  {Vielva}, {Villa}, {Vittorio}, {Wandelt}, {Wehus}, {White}, {White},
  {Zacchei}, \& {Zonca}}]{Planck_Collaboration_2020}
{Planck Collaboration}, {Aghanim}, N., {Akrami}, Y., {et~al.} 2020,
  \bibinfo{title}{{Planck 2018 results. VI. Cosmological parameters},} \aap,
  641, A6, \dodoi{10.1051/0004-6361/201833910}

\bibitem[{G. Pruto {et~al.}(2025)Pruto, Keating, Kannan, Puchwein, Smith,
  Borrow, Garaldi, Vogelsberger, Zier, McClymont, Shen, \&
  Tacchella}]{Pruto_2025}
Pruto, G., Keating, L., Kannan, R., {et~al.} 2025, \bibinfo{title}{The
  THESAN-ZOOM project: The Hidden Neighbours of OI Absorbers during
  Reionization,} \doarXiv{2510.13977}

\bibitem[{P. {Richter} {et~al.}(2013){Richter}, {Fox}, {Wakker}, {Lehner},
  {Howk}, {Bland-Hawthorn}, {Ben Bekhti}, \& {Fechner}}]{Richter_2013}
{Richter}, P., {Fox}, A.~J., {Wakker}, B.~P., {et~al.} 2013,
  \bibinfo{title}{{The COS/UVES Absorption Survey of the Magellanic Stream. II.
  Evidence for a Complex Enrichment History of the Stream from the Fairall 9
  Sightline},} \apj, 772, 111, \dodoi{10.1088/0004-637X/772/2/111}

\bibitem[{A.~M. Sebastian {et~al.}(2023)Sebastian, Ryan-Weber, Davies, Becker,
  Keating, D'Odorico, Meyer, Bosman, Cupani, Kulkarni, Haehnelt, Lai, Eilers,
  Bischetti, \& Gallerani}]{sebastian2023}
Sebastian, A.~M., Ryan-Weber, E., Davies, R.~L., {et~al.} 2023,
  \bibinfo{title}{E-XQR-30: The evolution of MgII, CII and OI across 2<z<6,}
  \doarXiv{2403.10072}

\bibitem[{A. {Sodini} {et~al.}(2024){Sodini}, {D'Odorico}, {Salvadori},
  {Vanni}, {Bischetti}, {Cupani}, {Davies}, {Becker}, {Ba{\~n}ados}, {Bosman},
  {Davies}, {Paolo Farina}, {Ferrara}, {Keating}, {Kulkarni}, {Lai},
  {Ryan-Weber}, {Maria Sebastian}, \& {Walter}}]{Sodini_2024}
{Sodini}, A., {D'Odorico}, V., {Salvadori}, S., {et~al.} 2024,
  \bibinfo{title}{{Evidence of Pop III stars' chemical signature in neutral gas
  at z {\ensuremath{\sim}} 6. A study based on the E-XQR-30 spectroscopic
  sample},} \aap, 687, A314, \dodoi{10.1051/0004-6361/202349062}

\bibitem[{L.~J. {Tacconi} {et~al.}(2018){Tacconi}, {Genzel}, {Saintonge},
  {Combes}, {Garc{\'\i}a-Burillo}, {Neri}, {Bolatto}, {Contini}, {F{\"o}rster
  Schreiber}, {Lilly}, {Lutz}, {Wuyts}, {Accurso}, {Boissier}, {Boone},
  {Bouch{\'e}}, {Bournaud}, {Burkert}, {Carollo}, {Cooper}, {Cox}, {Feruglio},
  {Freundlich}, {Herrera-Camus}, {Juneau}, {Lippa}, {Naab}, {Renzini},
  {Salome}, {Sternberg}, {Tadaki}, {{\"U}bler}, {Walter}, {Weiner}, \&
  {Weiss}}]{Tacconi2018}
{Tacconi}, L.~J., {Genzel}, R., {Saintonge}, A., {et~al.} 2018,
  \bibinfo{title}{{PHIBSS: Unified Scaling Relations of Gas Depletion Time and
  Molecular Gas Fractions},} \apj, 853, 179, \dodoi{10.3847/1538-4357/aaa4b4}

\bibitem[{T.~X. {Thuan} \& Y.~I. {Izotov}(1997){Thuan} \&
  {Izotov}}]{Thuan_1997}
{Thuan}, T.~X., \& {Izotov}, Y.~I. 1997, \bibinfo{title}{{Nearby Young Dwarf
  Galaxies: Primordial Gas and Ly{\ensuremath{\alpha}} Emission},} \apj, 489,
  623, \dodoi{10.1086/304826}

\bibitem[{L. {Tornatore} {et~al.}(2007){Tornatore}, {Ferrara}, \&
  {Schneider}}]{Tornatore_2007}
{Tornatore}, L., {Ferrara}, A., \& {Schneider}, R. 2007,
  \bibinfo{title}{{Population III stars: hidden or disappeared?},} \mnras, 382,
  945, \dodoi{10.1111/j.1365-2966.2007.12215.x}

\bibitem[{J. {Tumlinson} {et~al.}(2017){Tumlinson}, {Peeples}, \&
  {Werk}}]{tumlinson2017}
{Tumlinson}, J., {Peeples}, M.~S., \& {Werk}, J.~K. 2017, \bibinfo{title}{{The
  Circumgalactic Medium},} \araa, 55, 389,
  \dodoi{10.1146/annurev-astro-091916-055240}

\bibitem[{J. {Tumlinson} {et~al.}(2004){Tumlinson}, {Venkatesan}, \&
  {Shull}}]{Tumlinson2004}
{Tumlinson}, J., {Venkatesan}, A., \& {Shull}, J.~M. 2004,
  \bibinfo{title}{{Nucleosynthesis, Reionization, and the Mass Function of the
  First Stars},} \apj, 612, 602, \dodoi{10.1086/422571}

\bibitem[{J. {Tumlinson} {et~al.}(2011){Tumlinson}, {Thom}, {Werk},
  {Prochaska}, {Tripp}, {Weinberg}, {Peeples}, {O'Meara}, {Oppenheimer},
  {Meiring}, {Katz}, {Dav{\'e}}, {Ford}, \& {Sembach}}]{Tumlinson2011}
{Tumlinson}, J., {Thom}, C., {Werk}, J.~K., {et~al.} 2011, \bibinfo{title}{{The
  Large, Oxygen-Rich Halos of Star-Forming Galaxies Are a Major Reservoir of
  Galactic Metals},} Science, 334, 948, \dodoi{10.1126/science.1209840}

\bibitem[{D. {{\v{D}}urov{\v{c}}{\'\i}kov{\'a}}
  {et~al.}(2025){{\v{D}}urov{\v{c}}{\'\i}kov{\'a}}, {Eilers}, {Meyer},
  {Farina}, {Ba{\~n}ados}, {Davies}, {Hennawi}, {Mazzucchelli}, {Simcoe}, \&
  {Walter}}]{durovica2025}
{{\v{D}}urov{\v{c}}{\'\i}kov{\'a}}, D., {Eilers}, A.-C., {Meyer}, R.~A.,
  {et~al.} 2025, \bibinfo{title}{{Quasar Lifetime Measurements from Extended
  Ly{\ensuremath{\alpha}} Nebulae at z {\ensuremath{\sim}} 6},} \apj, 990, 174,
  \dodoi{10.3847/1538-4357/adf6dd}

\bibitem[{A. Venditti {et~al.}(2023)Venditti, Graziani, Schneider, Pentericci,
  Di Cesare, Maio, \& Omukai}]{Venditti_2023}
Venditti, A., Graziani, L., Schneider, R., {et~al.} 2023, \bibinfo{title}{A
  needle in a haystack? Catching Population III stars in the epoch of
  reionization: I. Population III star-forming environments,} Monthly Notices
  of the Royal Astronomical Society, 522, 3809–3830,
  \dodoi{10.1093/mnras/stad1201}

\bibitem[{F. {Wang} {et~al.}(2023){Wang}, {Yang}, {Hennawi}, {Fan}, {Sun},
  {Champagne}, {Costa}, {Habouzit}, {Endsley}, {Li}, {Lin}, {Meyer},
  {Schindler}, {Wu}, {Ba{\~n}ados}, {Barth}, {Bhowmick}, {Bieri}, {Blecha},
  {Bosman}, {Cai}, {Colina}, {Connor}, {Davies}, {Decarli}, {De Rosa}, {Drake},
  {Egami}, {Eilers}, {Evans}, {Farina}, {Haiman}, {Jiang}, {Jin}, {Jun},
  {Kakiichi}, {Khusanova}, {Kulkarni}, {Li}, {Liu}, {Loiacono}, {Lupi},
  {Mazzucchelli}, {Onoue}, {Pudoka}, {Rojas-Ruiz}, {Shen}, {Strauss}, {Tee},
  {Trakhtenbrot}, {Trebitsch}, {Venemans}, {Volonteri}, {Walter}, {Xie}, {Yue},
  {Zhang}, {Zhang}, \& {Zou}}]{ASPIRE_Wang_2023}
{Wang}, F., {Yang}, J., {Hennawi}, J.~F., {et~al.} 2023, \bibinfo{title}{{A
  SPectroscopic Survey of Biased Halos in the Reionization Era (ASPIRE): JWST
  Reveals a Filamentary Structure around a z = 6.61 Quasar},} \apjl, 951, L4,
  \dodoi{10.3847/2041-8213/accd6f}

\bibitem[{J.~K. {Werk} {et~al.}(2013){Werk}, {Prochaska}, {Thom}, {Tumlinson},
  {Tripp}, {O'Meara}, \& {Peeples}}]{Werk2013}
{Werk}, J.~K., {Prochaska}, J.~X., {Thom}, C., {et~al.} 2013,
  \bibinfo{title}{{The COS-Halos Survey: An Empirical Description of Metal-line
  Absorption in the Low-redshift Circumgalactic Medium},} \apjs, 204, 17,
  \dodoi{10.1088/0067-0049/204/2/17}

\bibitem[{O. Zier {et~al.}(2025)Zier, Kannan, Smith, Puchwein, Vogelsberger,
  Borrow, Garaldi, Keating, McClymont, Shen, \& Hernquist}]{Zier_2025}
Zier, O., Kannan, R., Smith, A., {et~al.} 2025, \bibinfo{title}{The THESAN-ZOOM
  project: Population III star formation continues until the end of
  reionization,} \doarXiv{2503.03806}

\bibitem[{E. {Zinger} {et~al.}(2020){Zinger}, {Pillepich}, {Nelson},
  {Weinberger}, {Pakmor}, {Springel}, {Hernquist}, {Marinacci}, \&
  {Vogelsberger}}]{Zinger_2020}
{Zinger}, E., {Pillepich}, A., {Nelson}, D., {et~al.} 2020,
  \bibinfo{title}{{Ejective and preventative: the IllustrisTNG black hole
  feedback and its effects on the thermodynamics of the gas within and around
  galaxies},} \mnras, 499, 768, \dodoi{10.1093/mnras/staa2607}

\end{thebibliography}
\bibliographystyle{aasjournalv7}

\section{Appendix} \label{sec:appendix}

In this appendix we present the four of the five \oi\ and \siii\ absorption lines associated with galaxy overdensities at $z\sim$6 presented in this work. The corresponding Voigt profiles and the total \oi\ column density of the systems are noted. Associated galaxies are marked as green ticks. The fifth system is presented in Figure \ref{fig:voigt_subplot_J0100_7406_XShooter_FIRE_HIRES_OI}.
 
\end{document}